\documentclass[usenatbib]{mn2e}
\usepackage{color} 
\usepackage{graphics}
\usepackage{epsfig}
\usepackage{natbib}

\begin{document}

\title[Galaxies with excess massive stars] 
{A resolved study of the inner regions of nearby galaxies with
an excess of young massive stars:  
missing link in the AGN-starburst connection?}                              

\author [G.Kauffmann] {Guinevere Kauffmann$^1$\thanks{E-mail: gamk@mpa-garching.mpg.de}, 
Iker Mill\'an-Irigoyen$^2$, Paul Crowther$^3$, Claudia Maraston$^4$ \\
$^1$Max-Planck Institut f\"{u}r Astrophysik, 85741 Garching, Germany\\
$^2$Departamento de Investigacion Basica, CIEMAT, Av. Complutense 40, E-28040, Madrid, Spain\\
$^3$Department of Physics and Astronomy, University of Sheffield, Sheffield S3 7RH, UK\\
$^4$Institute of Cosmology, University of Portsmouth, Burnaby Road, Portsmouth PO1 3FX, UK}

\maketitle

\begin{abstract} 
We have selected galaxies with very high levels of H$\alpha$ 
emission (EQW(H$\alpha$)$>$700 \AA.) in their central regions from the final data  release of the MaNGA survey . 
Our study focuses on 14 very well-resolved nearby galaxies with stellar masses
in the range $9.5 < \log M_*/(M_{\odot}) < 11.5$. 
We investigate a variety of  procedures for selecting galaxy regions that
are likely to harbour excess populations of young massive stars, finding that
selection in the 2-dimensional space of extinction-corrected H$\alpha$ EQW and
[SIII]/[SII] line ratio produces the best results. By  comparing
stacked spectra covering these regions with stacked  spectra covering 
normal starburst regions with 100\AA$<$EQW(H$\alpha$)$<$200\AA, we obtain the following main results: 
1) Clear signatures of excess  Wolf Rayet stars are found in half of the 
H$\alpha$ excess regions, 
2)Galaxy regions containing excess Wolf Rayet stars are  more  often
associated with the presence of  high-ionization emission lines 
characteristic of accreting black holes. Excess [NeIII] is detected in 4 out of 8 of the WR regions 
and there are tentative [FeX] detections in 2 galaxies.
3) Regions of the galaxy with excess  Wolf Rayet stars 
are located where the interstellar medium has larger ionized gas turbulent velocities
and higher  neutral gas overdensities.
We make a first attempt to constrain changes in the high mass end of
the stellar initial mass function (IMF)  using the HR-pyPopStar evolutionary
population synthesis models that include high wavelength-resolution theoretical
atmosphere libraries for Wolf Rayet stars.
\end {abstract}
\begin{keywords} galaxies: nuclei, galaxies: star formation, galaxies: stellar content,
stars: Wolf Rayet, galaxies: bulges, galaxies: active        
\end{keywords}

\section{Introduction}

The connection between active galactic nuclei (AGN) and starbursts has
been a vigorous  topic of research since the 1980's. Initially, much of the
debate in the field was focused on whether certain classes of AGN could be
entirely powered by star formation rather than by accreting black holes.
Terlevich \& Melnick (1985) proposed that because the most massive stars
can reach very high effective temperatures in the last stages of their
evolution, ionization by a population of extreme WC or WO Wolf-Rayet stars
could explain the observed luminosities of high ionization lines such as
[OIII]$\lambda$5007 and [NeIII]$\lambda$3869 in Seyfert 2 galaxies, and there was
therefore no need to invoke an accreting supermassive black hole as the main
powering source in these systems.

Subsequent searches uncovered a handful of Seyfert 2 galaxies with very
strong Wolf Rayet star signatures, such as Markarian 477 (Heckman et al 1997),
but the spectroscopic signatures of young stellar populations in most
objects tended to be much weaker (Gonzalez Delgado et al 1998,
Cid Fernandes et al 2001, Ho  et al 1997) and were generally in the form
of large far-IR luminosities, high near-UV surface brightnesses and emission
line ratios intermediate between pure Seyfert 2 galaxies and starbursts. As
a result, the Seyfert galaxy unification scheme introduced by Antonucci
(1993) in which viewing angle was the main factor in whether an AGN had a
Type I and Type II spectrum, became widely accepted and the Wolf-Rayet star
scenario dropped out of fashion.

The next insight into the role of star formation in the AGN phenomenon was
associated with the era of large surveys of galaxies, such as the Sloan
Digital Sky Survey (York et al 2000), where AGN could be selected using optical
line ratio diagnostic diagrams (Kauffmann et al 2003, Kewley et al 2006),
as well as large samples of X-ray and IR-selected AGN in the redshift range
0.2-2.5 (e.g. Mendez et al 2013; Stern et al 2015; Brown et al 2019). Heckman \& Kauffmann (2004)
estimated the volume-averaged ratio of star formation to black hole mass
accretion in galaxies and found that for early type galaxies with black hole
masses in the range $10^7-10^{8.5} M_{\odot}$, this value was around 1000,
in good agreement with the empirically-derived ratio of bulge mass to black
hole mass in nearby galaxies (Marconi \& Hunt 2003; H\"aring \& Rix 2004). This result appeared to
suggest a close relationship between star formation in the bulge and the
fuelling/growth of supermassive black holes.

Later work showed that this relationship  derived by integrating the total
star formation and black hole growth rates over large populations of objects,
does not extend in a simple way to individual systems. Kauffmann \& Heckman
(2009) analyzed the observed distribution of Eddington ratios in galaxies
split into bins of black hole mass and showed that in objects with young
stellar populations, the Eddington ratio distribution was characterized by a
broad lognormal distribution of accretion rates peaked at a few per cent of
the Eddington limit. Moreover, the distribution depended very little on the
central stellar population of the galaxy, as characterized by its 4000 \AA\
break strength. In this work, the Eddington ratios were estimated using
the [OIII] line luminosity. A subsequent analysis used X-ray luminosity
as a black hole accretion rate indicator and IR luminosity to measure star
formation rate, and reached very similar conclusions (Stanley et al 2015). A
reservoir of gas in the host galaxy is clearly necessary for both ongoing
star formation and black hole growth, but does not closely regulate the latter.

Attention in the field has now turned back to asking whether there are
particular interstellar medium conditions or configurations that are favourable
to the initial formation of a supermassive black hole in a galaxy or a major
change in its mass over a short time interval.  The compression of gas during
the merging of galaxies has long been hypothesized to be an important mechanism
to fuel the brightest AGN in the Universe (Sanders et al 1988).  Once again,
a clear link between AGN and merging galaxies has not held up  under careful
statistical investigation (e.g. Li et al 2008; Reichard et al 2009; Kocevski
et al 2012). On the other hand, mergers are not the only mechanism that cause
gas compression in galaxies.  Black hole accretion is a process  that
occurs on very small scales and it is likely that local interstellar medium
conditions need to be measured to find the clearest AGN-starburst connections.

The  Sloan Digital Sky Survey IV included a survey called MaNGA (Mapping
Nearby Galaxies at APO), which obtained spectral measurements across the face
of each of $\sim$10,000 nearby galaxies thanks to 17 simultaneous integral
field units (IFUs) (Bundy et al 2015, Yan et al 2016). 
In a number of papers, AGN have been identified using
neubular emission line ratio diagnostics (Baldwin, Phillips \& Terlevich 1981;
hereafter BPT)
 and their resolved star formation
profiles have been studied. The main conclusion from these studies is that
BPT-selected AGN have suppressed central star formation rates compared to
control samples matched in stellar mass and redshift (Guo et al 2019; Jin et
al 2021; Lammers et al 2022), leading to the conclusion that AGN have acted to
quench star formation in galaxies from "inside-out". Other studies have focused
on searching for "hidden" populations of  accreting black holes in more
strongly star-forming galaxies using weaker, but higher-ionization emission
lines such as [NeV]$\lambda$3427 (Negus et al 2023) or HeII$\lambda$4686
(Tozzi et al 2023). These galaxies are hypothesized to be at an earlier
stage of an accretion event onto the black hole, when active star-formation
is still ongoing.

Our previous work has focused on the hypothesis that black holes may form and accrete
in environments that are unusually rich in  very young, massive stars.
We initially focused   on selecting galaxies with extinction-corrected
H$\alpha$ equivalent widths too large to be explained with a standard initial
mass function. In a study of these galaxies in an early data release of
the SDSS-IV MaNGA survey (Aguado et al 2019), 
Kauffmann (2021) found that the 4000 \AA\ break
is either flat or rising towards the centre of the galaxy, indicating that
the central regions host evolved stars, but the H$\alpha$ equivalent width
also rises steeply in the central regions, a trend that cannot be explained
by any star formation history assuming a standard IMF. The  implication
is that  there is an  excess of 
ionizing  sources near the centres of these galaxies.

 Subsequent work (Kauffmann et al 2022)  focused on clarifying the nature of the ionizing sources
in very high  H$\alpha$ equivalent width galaxies in a much larger sample        
selected from the SDSS
main sample with single-fibre rather than IFU spectra. This study found [Ne V]
emission in a subset of the H$\alpha$ excess galaxies with radio detections
indicating that accreting black holes were likely present in a subet of the objects.
It also revealed strong and broadened He II$\lambda$4686 emission lines
characteristic of stars from the  WN sequence  Wolf Rayet stars in some of the galaxies.  Because the signal-to-noise
of the single fibre SDSS spectra is low compared to MaNGA (1 hour exposures
compared to 3 hour exposures on the APO 2.5 metre telescope), the analysis
could only be carried out for stacked spectra of a few dozen galaxies. 
Even so, the stacked  spectra were  too low in $S/N$ to detect the high-ionization
 (CIII and CIV) lines associated with WC sequence Wolf Rayet stars.   
The study could also not focus on local interstellar medium conditions within
individual galaxies.

In this paper, we return to samples of H$\alpha$ excess galaxies selected
from the final release of the MaNGA survey (Abdurro'uf et al 2022). 
Our study focuses on well-resolved nearby galaxies with stellar masses
in the range $9.5 < \log M_*/(M_{\odot}) < 11.5$, i.e. within  a factor of 10-20 the 
mass of our own Milky Way, which is estimated as $6\times10^{10} M_{\odot}$
(Licquia \& Newman 2015). We refine our  procedures for selecting galaxy regions that
are likely to harbour excess populations of young massive stars. We compare
stacked spaxel spectra of these regions with stacked spaxel spectra of more
normal star-forming regions within the same galaxy. 
Finally, we make a first attempt to constrain changes in the high mass end of
the stellar initial mass function (IMF)  using the HR-pyPopStar evolutionary
population synthesis models that include high wavelength-resolution theoretical
atmosphere libraries for Wolf Rayet stars (Mill\'an-Irigoyen et al 2021).

Our paper is organized as follows. In section 2, we describe how the galaxy
sample analyzed in this paper is selected. In section 3, we discuss our
spaxel selection and stacking procedure. Section 4 presents two-dimensional
maps of physical properties derived from individual spaxels. Sections 5 and
6 present results on Wolf-Rayet features and high-ionization emission
lines, while section 7 focuses on interstellar medium properties. Section 8
explores the effect of IMF changes predicted by the HR-pyPopStar models. A
summary and discussion of the main results is presented in section 9.

\section {The parent galaxy sample}
The galaxies are selected from the MaNGA DRPALL file, which provides basic photometric
information and physical quantities such as stellar mass for the  10,010  unique  galaxies 
with data cubes available as part of the final MaNGA data release (DR17). Most of the
galaxy properties in the files are taken from the NASA-Sloan Atlas (NSA) catalogue,
a catalog of images and parameters of local galaxies derived from Sloan Digital Sky Survey imaging.
The object detection, deblending, and other details regarding the image analysis can
 be found in Blanton et al. (2011). The image analysis is better tuned for large, 
bright galaxies than the standard SDSS processing. 

 We apply a cut on
the ratio of the galaxy major-to-minor axis lengths $b/a > 0.75$ to select face-on systems, a cut
on stellar mass $9.5 < \log M_*/(M_{\odot}) < 11.5$ to select galaxies within a factor of 10 the
mass of our own Milky Way, and a redshift cut $0.005<z<0.05$ to select the most nearby
systems with better-resolved coverage of the  inner stellar population.  This procedure yields  1858 data cubes 
that are listed as having high quality data.

The MaNGA data-analysis pipeline (MaNGA DAP, Westfall et al 2019) is the 
survey-led software package that analyzes the data 
produced by the MaNGA data-reduction pipeline (MaNGA DRP, Law et al 2016) 
to produce physical properties derived from 
MaNGA spectroscopy. The DAP output primarily consists of two output files, the MAPS and model LOGCUBE files, 
provided for each combination of PLATE-IFU and DAPTYPE. The MAPS file provides 2D "maps" (i.e., images) of 
DAP measured properties, such as the fluxes and equivalent widths of emission lines, absorption line
spectral indices and stellar kinematic quantities such as $V$ and $\sigma$.
We note that  the instrumental spectral line-spread function (LSF) typically has a 1$\sigma$ width of about 70 km/s
(Law et al 2021).
To select galaxies with excess H$\alpha$ emission, we make use of the H$\alpha$ and H$\beta$ flux
and continuum measurement maps provided by the HYB10-MILESHC-MASTARSSP MAP files. In these files,
the stellar-continuum measurements are made from  spectra binned to S/N $\sim 10$ using a Voronoi binning algorithm,
but the emission-line measurements are performed on the individual spaxels, yielding the highest
possible spatial resolution for these measurements.  

We use the MAPS files to calculate the average extinction-corrected H$\alpha$ equivalent width in 
10 equally spaced radial
bins from 0.025 to 0.5 R$_{50}$, where R$_{50}$ is the r-band half-light radius provided by the DRPALL file. 
We adopt the same procedure for extinction correction using the Balmer decrement H$\alpha$/H$\beta$ 
described in Kauffmann (2021) and we retain 58 galaxies
where at least one radial bin has an average H$\alpha$ EQW greater than 700, placing it into the category
of H$\alpha$ excess systems, where the extinction-corrected line strength is a factor of two
higher than that predicted for
a starburst of zero age  with a standard Kroupa (2001) or Chabrier (2003)   initial mass function
(see Kauffmann (2021) for more details).
We select 17 galaxies with the best spatially resolved observations, defined as the number of
spaxels within the very central region of the galaxy ($R< 0.15 R_{50})$. 

Table 1 lists some of  the key properties of the galaxies that form part of the study in this paper.
These include the MaNGA identifier, position, redshift, NUV-r colour, half-light radius $R_{50}$, 
axial ratio ($b/a$), Sersic index and stellar mass. In addition, we list the number of spaxels
included within 0.15$R_{50}$ and 0.5$R_{50}$. 
Physical properties are listed  for the sample with 
clearly detected Wolf-Rayet features  and the sample with no detected features in separate sections of
the table.  The three galaxies that are are not listed in the table  are classified as weak detections and do
not form part of the analysis. The Wolf Rayet feature detection procedure
and classification will be discussed in more detail in the next section.

We note the galaxies in this sample have $R_{50}$ values between 5 and 20 arcsec.
Table 1 shows that for the selected sample,  the region $R< 0.5R_{50}$
is typically covered  by  200-300 spaxels and the very central region of the galaxy 
 $R< 0.15R_{50}$ is covered by 10-30 spaxels.  As we will show, stacking of subsets of  
very high H$\alpha$ EQW spaxel spectra 
in the galaxy and comparing these spectra with stacked spectra of spaxels in ordinary
starburst regions is useful for  
pulling out even relatively weak Wolf-Rayet features.  

All the galaxies in both the  Wolf-Rayet detected and non-detected sub-samples 
 have stellar masses in the range $\log M_*/(M_{\odot})=10-10.6$.
Kauffmann et al (2022) extracted a sample of 1000 H$\alpha$ excess galaxies from
the SDSS main sample with single fibre spectra and  showed that the 
stellar mass distribution of galaxies with H$\alpha$ excess
peaks in the range $10^{10} -10^{10.5} M_{\odot}$.  

With one exception, the Sersic indices in the Wolf-Rayet sample are below 2, indicating that
they are late-type spiral systems. Four out of six of the  galaxies without Wolf-Rayet feature detections
have Sersic indices greater than 2, indicating the stellar light in these systems  is more concentrated.
We note that the Kauffmann et al (2022) study only detected Wolf-Rayet features in stacked spectra
of many galaxies rather than in individual galaxies, so it was not possible to investigate systematic changes in 
galaxy morphology.

Figures 1 and 2 present SDSS g,r,i-band postage stamp images of the two 
sub-samples with and without  Wolf-Rayet detections.
In addition, we also show VLA FIRST 1.4 GHz cut-out images of all
the sources.  Both sub-samples contain one galaxy merger (9507.12704 and 11830.12701). The Wolf-Rayet
detected sub-sample contains 2 galaxies with close companions (7959.6103 and 8311.6104). Both sub-samples also
contain galaxies where there is no clear perturber. In the cases where there is a high $S/N$ radio detection,
the radio emission appears to be unresolved in most cases. Occasionally, the radio emission is significantly
offset from the center of the galaxy (e.g. 8336.12704). At the resolution of the VLA images, the   
radio properties or evidence of mergers/interactions  do not provide a clear 
predictor of whether or not  Wolf-Rayet
features will be detected in the spectra.

\begin{figure*}
\includegraphics[width=171mm]{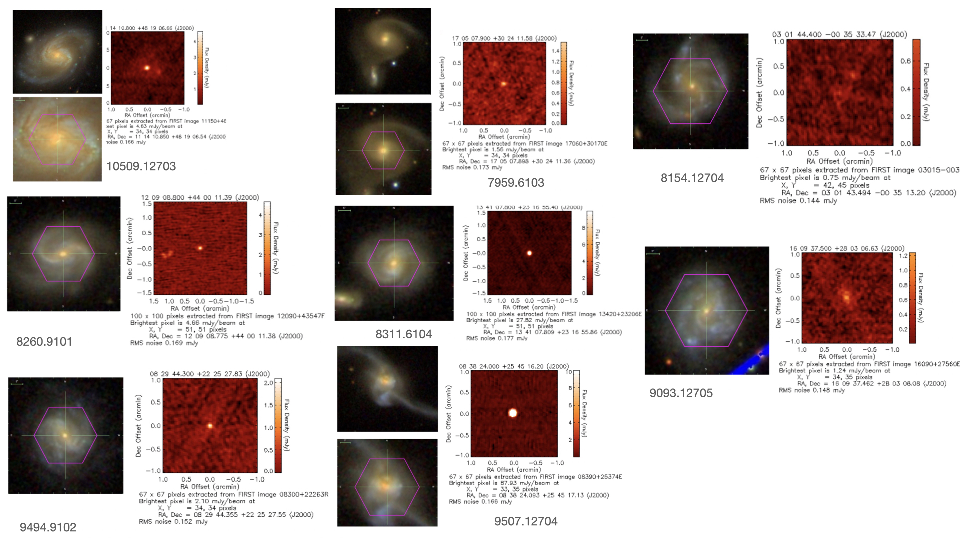}
\caption{ A gallery of SDSS $g,r,i$ and VLA FIRST 1.4 GHz postage stamp
images of the 8 galaxies with Wolf Rayet feature detections.
\label{models}}
\end{figure*}

\begin{figure*}
\includegraphics[width=171mm]{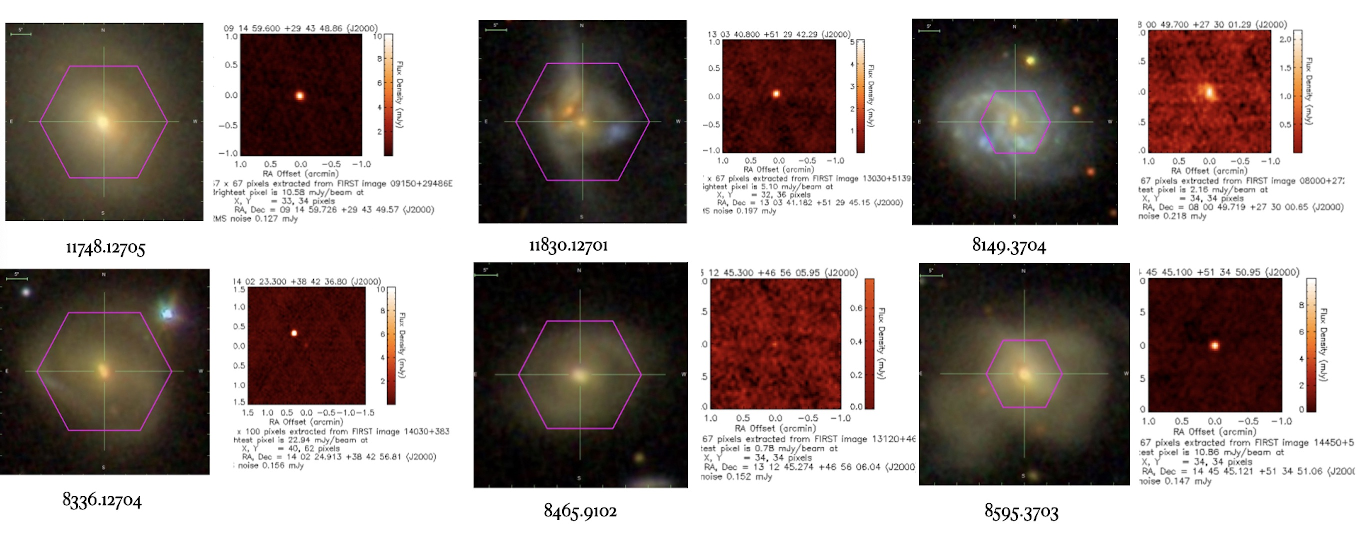}
\caption{ A gallery of SDSS $g,r,i$ and VLA FIRST 1.4 GHz postage stamp
images of the 6 galaxies with no Wolf Rayet feature detections.
\label{models}}
\end{figure*}

\begin{table*}
\caption{Table of physical quantities for the H$\alpha$ excess samples
studied in this paper (top: 8 galaxies with WR detections, bottom: 6
galaxies with no WR detections)\\
The columns are as
follows: 1) MaNGA identifier (plate ID), 2) MaNGA identifier (IFU design number), 
3) right ascension (J2000), 4) declination (J2000), 5) redshift,
6) NUV-r colour, 7) r-band half-light radius (arcsec) ,
8) axial ratio b/a, 9) Sersic index  9) logarithm of the stellar mass ($M_{\odot}$),
10) number of spaxels within 0.15R$_{50}$, 11) number of spaxels within 0.5R$_{50}$}
\resizebox{\textwidth}{!}{%
\begin{tabular}{r|r|r|r|r|r|r|r|r|r|r|r}
\hline			
plate id & ifu design no. &  RA & DEC & z & NUV-r  & r50  & b/a  & Sersic n  & log (M$_*$) &N$_{spax}$(0.15R$_{50}$)& N$_{spax}$(0.5R$_{50}$) \\ \hline  
10509& 12703& 168.545376& 48.318516&  0.007122&  3.224800& 20.110100& 0.806780& 1.633650& 10.353810 & 113 & 1273\\  
 7959&  6103& 256.283191& 30.403216&  0.034064&  5.155600&  6.317760& 0.958269& 5.404750& 10.575510 & 29 & 325\\
 8154& 12701&  45.435110& -0.592630&  0.031465&  2.642700& 10.275900& 0.842325& 0.512912& 10.237617 & 30 & 335\\
 8260&  9101& 182.286727& 44.003165&  0.037542&  2.741300&  8.080410& 0.761730& 1.528640& 10.594678 & 21 & 241\\ 
 8311&  6104& 205.282731& 23.282055&  0.026353&  3.311900&  5.444330& 0.835267& 1.871530& 10.283634 & 9 & 97\\  
 9093& 12705& 242.406339& 28.051843&  0.032796&  2.225500&  7.815390& 0.843981& 1.068260& 10.116239 & 21 & 193\\  
 9494&  9102& 127.434839& 22.424396&  0.025005&  2.177800&  6.606660& 0.783767& 1.501930& 10.062650 & 9 & 137\\  
 9507& 12704& 129.600037& 25.754501&  0.018181&  3.152400&  7.900920& 0.850208& 1.847050& 10.185556 & 21 & 193\\  
\hline  
11748& 12705& 138.748621& 29.730240&  0.021161&  4.299219&  8.687565& 0.824277& 5.669246& 10.676188 & 21 & 241\\  
11830& 12701& 195.920019& 51.495081&  0.038110&  3.037155&  7.566992& 0.755167& 1.589030& 10.346302 & 21 & 277\\  
 8149&  3704& 120.207222& 27.500359&  0.017345&  2.116600&  7.544130& 0.843134& 0.790922&  9.994284 & 29 & 321\\  
 8336& 12704& 210.597154& 38.710223&  0.019987&  4.052200& 10.364800& 0.875560& 2.614080& 10.098990 & 29 & 341\\  
 8465&  9102& 198.189163& 46.934987&  0.027681&  3.901800&  7.458280& 0.837129& 3.513580& 10.061317 & 21 & 241\\  
 8595&  3703& 221.437984& 51.580820&  0.029627&  3.798600&  7.070540& 0.884235& 4.139190& 10.585714 & 13 & 153\\  
\hline
\end{tabular}}
\end{table*}

\section {Selection of Spaxels and Stacking Procedure}

Early attempts to find Wolf-Rayet galaxies relied on imaging strongly star-forming systems with a
narrowband $\lambda$4684 filter (Drissen et al 1990, Sargent \& Filippenko et al 1991).
Catalogues of Wolf-Rayet galaxies were compiled by a number of authors as more
and more objects were discovered (e.g. Conti 1991, Schaerer et al 1999,  Guseva et al 2000)
 In recent years, searching
for Wolf-Rayet features in spectra obtained as part of large galaxy  surveys has become the norm.
Catalogues constructed from single-fibre observations of galaxies from  the Sloan Digital
Sky Survey  have been constructed by first pre-selecting stronger emission line
objects based on their measured H$\beta$  (Brinchmann et al 2008) or  
H$\epsilon$ equivalent widths (Zhang et al 2007)  to select the strongest star-forming systems  and then searching for
a broad emission features (blue bump) in the vicinity of 4650 \AA. Such techniques were extended to
individual spaxel spectra in integral field unit (IFU) observations of galaxies  obtained as part of the 
Mapping Nearby Galaxies at APO (MaNGA) survey by Liang et al (2020). Miralles-Caballero et al (2016) 
identified extragalactic regions with Wolf-Rayet stars in IFU spectra obtained by the Calar Alto
Legacy Integral Field Area (CALIFA) survey (S\'anchez et al 2012). Their search algorithm  selected spaxels with
large H$\alpha$ equivalent widths, subtracting a best-fit continuum model and  then applying a pseudo-filter spanning the
wavelength range 4600-4700 \AA.  A similar procedure was applied to detect the red bump.   

In this paper, we introduce a spectrum stacking technique to increase  the signal-to-noise of the 
resulting spectra. Rather than searching for pre-defined specific features  in individual spaxel spectra,  we
proceed  by comparing stacked spectra in regions where ionized gas indicators
indicate that the local density of massive stars is unusually high with stacked spaxel spectra from
more normal regions within the same galaxy.
This technique has the advantage of increasing the  signal-to-noise in the spectra that
we analyze and enables us to look for weak features associated with other correlated phenomena, such
as emission from accreting black holes that are hidden within the starburst (Kauffmann et al  2022). The inner regions of strongly
star-forming galaxies also contain a lot of dust and as we will show, interstellar absorption features
are very strong and this can also hinder accurate characterization of the stellar continuum that
is necessary to dig out the weaker emission lines.   

Our technique  could have the disadvantage that because
the pre-selection is based on emission line properties, areas of the galaxy where Wolf-Rayet stars are
not correlated with a high density  of  ionized gas may be missed. Because the spatial resolution in MaNGA is quite low
and the regions that we are picking out for stacking typically extend over a contiguous
region  of a few hundred parsecs to a kiloparsec, we believe this is unlikely to be a major
problem. We speculate that this may be more of an issue in the outer disks of star-bursting galaxies, which
often contain isolated, compact clumps of very young stars embedded in a surroung lower-density interstellar medium.

We experimented with a variety of strategies for spaxel stacking with view to extracting 
visible Wolf-Rayet features in our central galaxy  spectra.
The strategies can be summarized as follows: 1) stack spectra with the highest extinction-corrected  
H$\alpha$ equivalent widths, 2) stack spectra selected both by extinction-corrected H$\alpha$ equivalent width and a measure
of the ionization state of the gas. In star-forming galaxies, ionizing photons are produced by the most massive
stars that are present and the ionization parameter provides an independent indicator of the local surface density of
young massive stars. 

There are a number of line ratios that are commonly used as diagnostics of the ionization state of
the photo-ionized gas in the interstellar medium of galaxies. [OIII]$\lambda$5007/[OII]$\lambda$3727
is strongly affected by dust reddening and is not considered in this work. [OIII]$\lambda$5007/H$\beta$
has the advantage that the lines are very closely spaced so that dust reddening effects are minimized, but
this ratio is strongly dependent on metallicity. In recent work, 
Sanders et al (2020) introduce the line ratio
S$_{32}$=[SIII]$\lambda\lambda$9069,9531/[SII]$\lambda\lambda$6716,6731 as an ionization diagnostic. Since
the lines are located at relatively long wavelengths, this ratio is not very sensitive to
dust and CLOUDY models (Ferland et al. 1998,2017)  indicate that it is even more weakly dependent on metallicity than
[OIII]$\lambda$5007/[OII]$\lambda$3727. 

By trying out different methodologies, we find that a spaxel cut in the 2 dimensional space of H$\alpha$ equivalent width
and ionization parameter is better at picking out Wolf-Rayet feratures than a cut on  H$\alpha$ equivalent width
alone. This probably indicates that the dust corrections that we apply to deredden EQW H$\alpha$  are not
fully accurate. We also find that cuts using  EQW H$\alpha$ and S$_{32}$ sometimes work better than cuts
on  EQW H$\alpha$ and [OIII]/H$\beta$. An example of such a case is shown in Figure 3. 
The upper right panel shows the spaxels selected for stacking in the plane of  EQW H$\alpha$ versus  [OIII]/H$\beta$.
The spectra from the spaxels with measurements falling  within the magenta box are combined to form the
spectra plotted in magenta in the middle and top-left  panels.  We normalize each spectrum to the average 
flux integrated over the wavelength range 5345-5410 \AA\, which is free of  emission lines
and strong absorption features.     
The spectra from the spaxels with measurements falling  
within the green box are stacked to show the spectra of ``normal starburst'' regions of the galaxy
for comparison. 
Comparison of the stacked spectra over the wavelength regions covering the HeII$\lambda$4686 
and CIV$\lambda$5801 emission lines are used to select galaxies with detectable  Wolf-Rayet excess. 

The top panels of Figure 3 show that for the galaxy 8260.9101 
there is marginally significant difference ($<2 \sigma$ in the region of the emission lines)  
between the two spectra if spaxels are selected using the [OIII]/H$\beta$ ratio. This system
would be classified as a weak detection.  
The bottom panels of Figure 3 show what happens if the same exercise carried out
in the plane of  EQW H$\alpha$ versus S$_{32}$.  The spaxel measurements  bifurcate
into two tracks in this plane -- one where there is no clear relation between  EQW H$\alpha$ and  S$_{32}$
and another track where there is a clear correlation. Stacking the high  EQW H$\alpha$ spaxels from this
second track now yields  spectra with clear excess  HeII$\lambda$4686 and CIV$\lambda$5801 emission,
so the galaxy is now classified as a detection.

In Figure 4, we show the spaxel selection boxes for all eight of the galaxies with Wolf-Rayet excesses.
The spaxel selection boxes for the high ionization stacks are chosen so that they  contain 1-2 dozen  spaxels with the
highest H$\alpha$ equivalent widths and the highest ionization parameters. The green selection boxes are
chosen to cover spaxels in the  H$\alpha$ equivalent width range 80-200 \AA\ with the lowest ionization parameters.
These values are  chosen because this is the expected value for a starburst galaxy with  a normal IMF (see for example Figure 10
in  Kauffmann (2021)).
In two cases (8311.6104 and 7959.6103), we are forced to change the spaxel selection boundaries  
because there only a  few  spaxels with EQW(H$\alpha$) in the range 80-200 \AA.

In four out of the eight cases, the  EQW H$\alpha$ versus S$_{32}$  cut yields stronger excess 
 HeII$\lambda$4686 and/or  CIV$\lambda$5801 emission and is adopted as our default. In the cases where
 [OIII]/H$\beta$ selection is adopted, S$_{32}$  almost always yields very similar (i.e. not worse)
results. The corresponding spectra in the blue and red bump wavelength regions are
presented in Figure 8 and 9 and are discussed in section 5.  We note that our Wolf-Rayet detection fraction of 
50\% in H$\alpha$ excess galaxies is very high compared
to what is obtained for the star-forming galaxy population as a 
whole.  Liang et al (2020)
derive  a Wolf-Rayet detection fraction in star-forming MaNGA galaxies that increases towards lower
stellar masses, peaking at a value of 12\%  at $\log M_* \sim 10^{9.75} M_{\odot}$.

\begin{figure*}
\includegraphics[width=161mm]{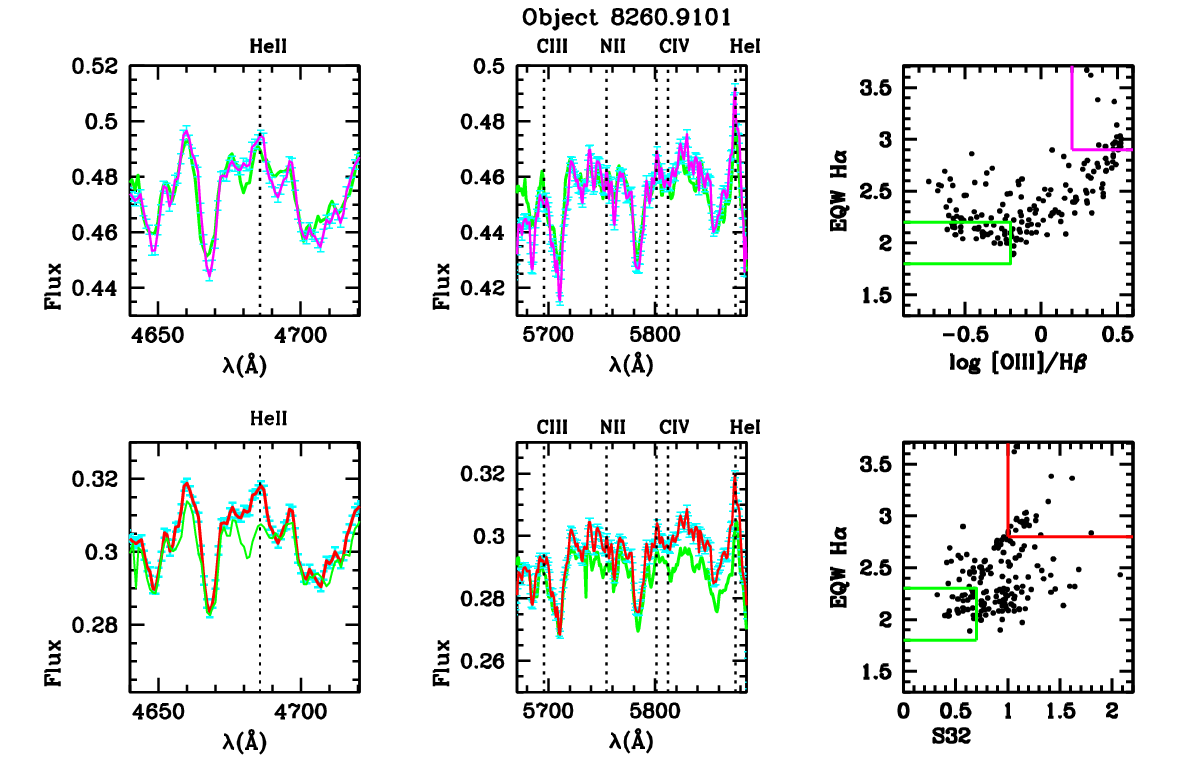}
\caption{ An example where selection of spaxels in the plane of  EQW H$\alpha$ versus
S32 (see bottom  panels) leads to stronger WR features than selection in the plane of  EQW H$\alpha$
versus  [OIII]/H$\beta$ (see top panels). The blue bump region is plotted in the left panels and the
red bump region in the right planels. The red,  magenta and green  lines show the stacked spectra
for the spaxels with emission line measurements located in the red,magenta and green bounding
boxes indicated in the right-hand panels (see explanation in text for bounding boxes). 
The spectra plotted in green indicate 
``normal starburst'' regions of the galaxy, whereas the spectra plotted in red and
magenta indicate regions with high H$\alpha$ equivalent width and high ionization.
The 1$\sigma$ measurement errors on the red and magenta stacked spectra are plotted in cyan.
\label{models}}
\end{figure*}

\begin{figure*}
\includegraphics[width=151mm]{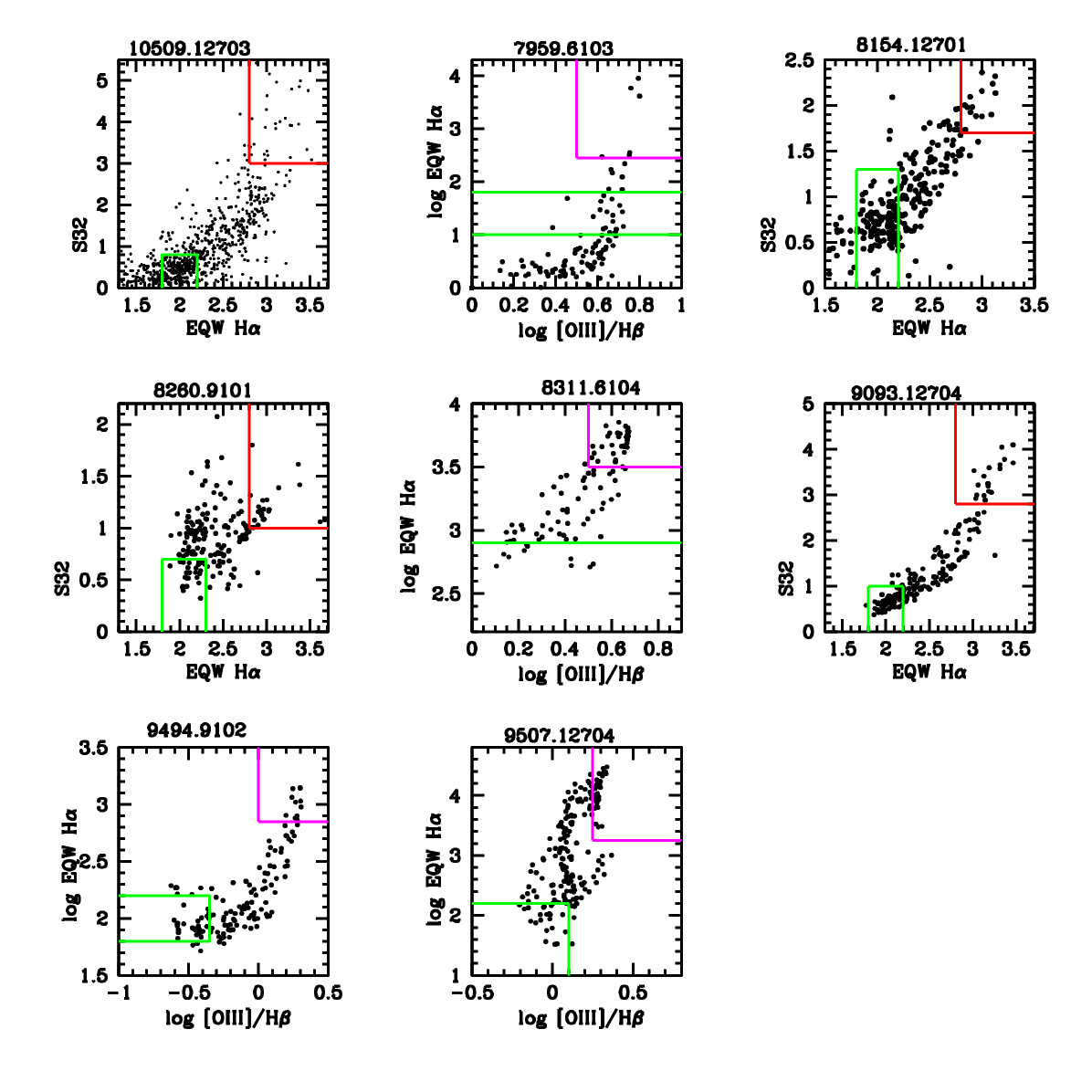}
\caption{The  selection of spaxels in the plane of  EQW H$\alpha$ versus
S32 or  EQW H$\alpha$
versus  [OIII]/H$\beta$ is shown for all 8  objects with WR feature detections.
\label{models}}
\end{figure*}

\section{Maps of spectral quantities}

Figures 5-8 present two-dimensional maps showing how the dust-corrected H$\alpha$ equivalent width and
ionization parameter [OIII]/H$\beta$
vary within the half-light radius R$_{50}$. The maps have all been scaled to the half-light radius
of each object (see caption of Figure 5 for more details). Results are shown separately for the 8 galaxies with
Wolf-Rayet excess detections and the 6 galaxies with no Wolf-Rayet excess. In Figure 5, we indicate the
spaxels that are included in the stacked spectra with Wolf-Rayet excess using large magenta triangles.
As can be seen, these spaxels are sometimes, but not always located near the very central regions
of the galaxy. In two  out of eight  galaxies, the stacked spectra include spaxels that extend  
as far as the $r$-band half-light radius R$_{50}$ and in two galaxies, the stacked spaxels do not include the very central regions.
The Wolf-Rayet emission is strongly offset
in objects 7959.6103, 9507.12704 and weakly offset in 10509.12703. In  objects
8154.12701, 8260.9101, and  9507.12704, it is exhibits a strongly asymmetric morphology 
with the spaxels aligned along a particular direction. In object 8154.12701, the
directional alighnment is so pronounced that one might hypothesize that the Wolf-Rayet
emitting regions are tracing an interstellar medium component that has been shocked
and compressed by a jet.   

Comparison of Figures 5 and 6 do not reveal clear differences. 
This means that the H$\alpha$ emission is distributed much
the same way in the two sub-samples with and without WR  features.
A bar is provided at the bottom-left of each panel
that can be used to calibrate  the physical size of the Wolf-Rayet regions under study.
The sizes range from 300pc to over 1 kpc.

Comparison of Figures 7 and 8 indicates that the 
galaxies with Wolf-Rayet excess emission usually have more highly ionized gas
within R$_{50}$.  In the Wolf-Rayet excess subsample, five out of eight galaxies have 
spaxels where $\log$ [OIII]/H$\beta$ $>0.4$ -- these are values that are often associated with active
galactic nuclei.  In the sub-sample with no Wolf-Rayet excess, only one galaxy out of six
has measured [OIII]/H$\beta$ values this high. We note that in both sub-samples the ionization parameter is usually
quite uniform with occasional patches of more highly ionized gas spread throughout the
central region within R$_{50}$. Only in two out
of eight galaxies in the WR excess sample is there a strong gradient in ionization (with higher
values towards the central regions). In the sample without WR excess, there are no galaxies  where
the ionization parameter peaks near the galaxy centre. This disfavours a centrally-located
active galactic nucleus as the main source of the ionization in most of these galaxies.

In the Appendix, we  show  two additional sets
of maps of two quantities: a) the Balmer absorption line feature H8, which is sensitive to
the age of the underlying population of older stars in the galaxy, b) ionized gas velocity
measured from the shift of the centroid of the H$\alpha$ emission line. No clear differences
in these maps are found for the two sub-samples. We do not consider ionized gas-phase metallicity
in this paper. As we will show, the H$\alpha$ excess sample with Wolf Rayet star detections 
often also exhibits signatures of ionization from accreting black holes, but the 
 H$\alpha$ excess sample without Wolf Rayet star signatures does not. This precludes
a simple comparison of ionized gas-phase metallicity between the two samples.

\begin{figure*}
\includegraphics[width=118mm]{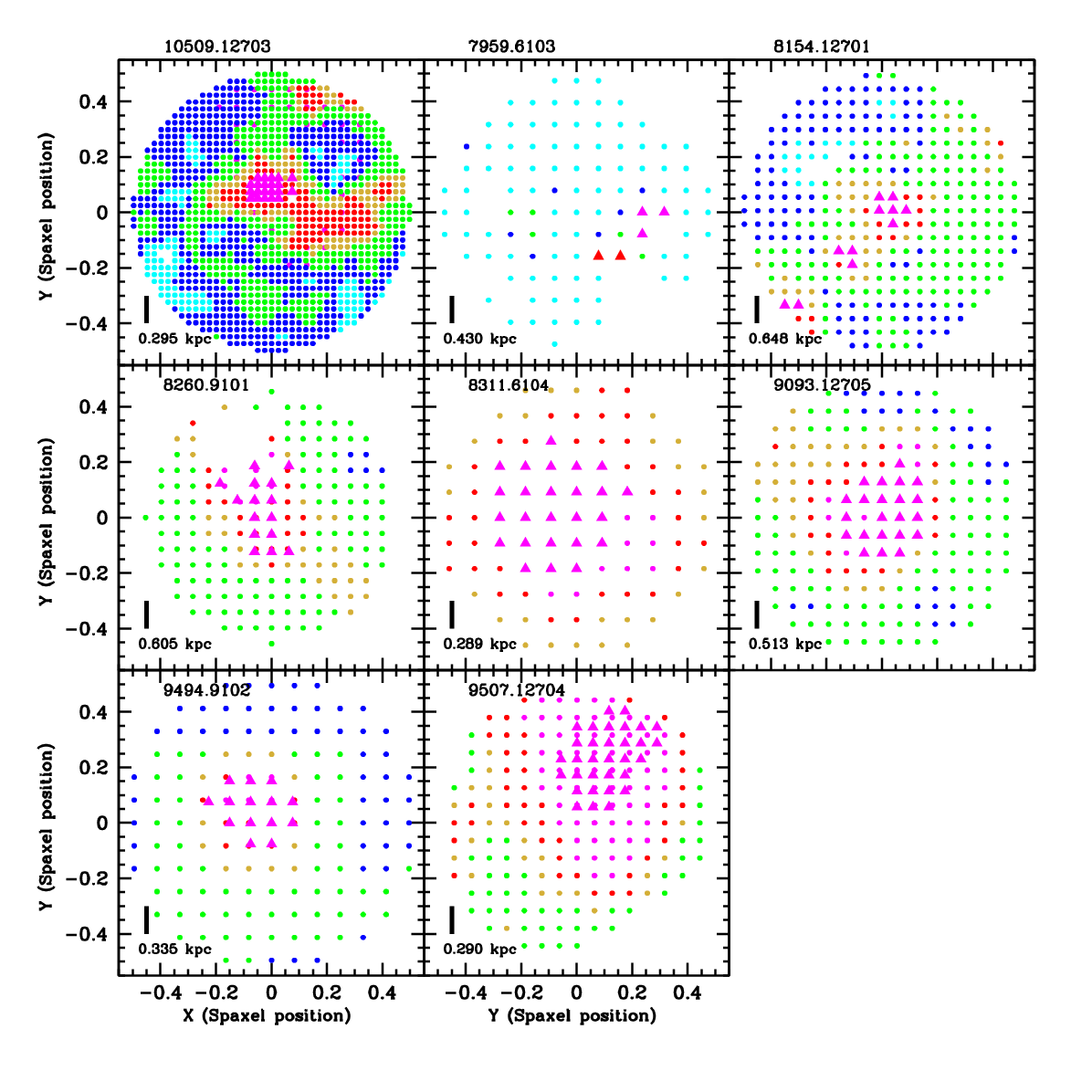}
\caption{Two-dimensional maps of H$\alpha$ equivalent width measurements for the 8 galaxies
with Wolf-Rayet detections. The circular points indicate each spaxel measurement and are
colour-coded as follows: magenta (EQW(H$\alpha$)$>$3.0), red (2.5$<$EQW(H$\alpha$)$<$3.0),  
dark gold (2.0$<$EQW(H$\alpha$)$<$2.5), blue(1.5$<$EQW(H$\alpha$)$<$2.0),
cyan(EQW(H$\alpha$)$<$1.5). The triangular points color-coded in magenta indicate those
spaxels that were stacked to produce the spectra with detectable WR excess.
The X and Y-axes are in units of the the scaled  radius measured
in the $r$-band, with a value 0.5 indicating the half-light radius.
A bar is provided at the bottom-left of each panel
that can be used to calibrate  the physical size of the regions under study.  
\label{models}}
\end{figure*}
\begin{figure*}
\includegraphics[width=121mm]{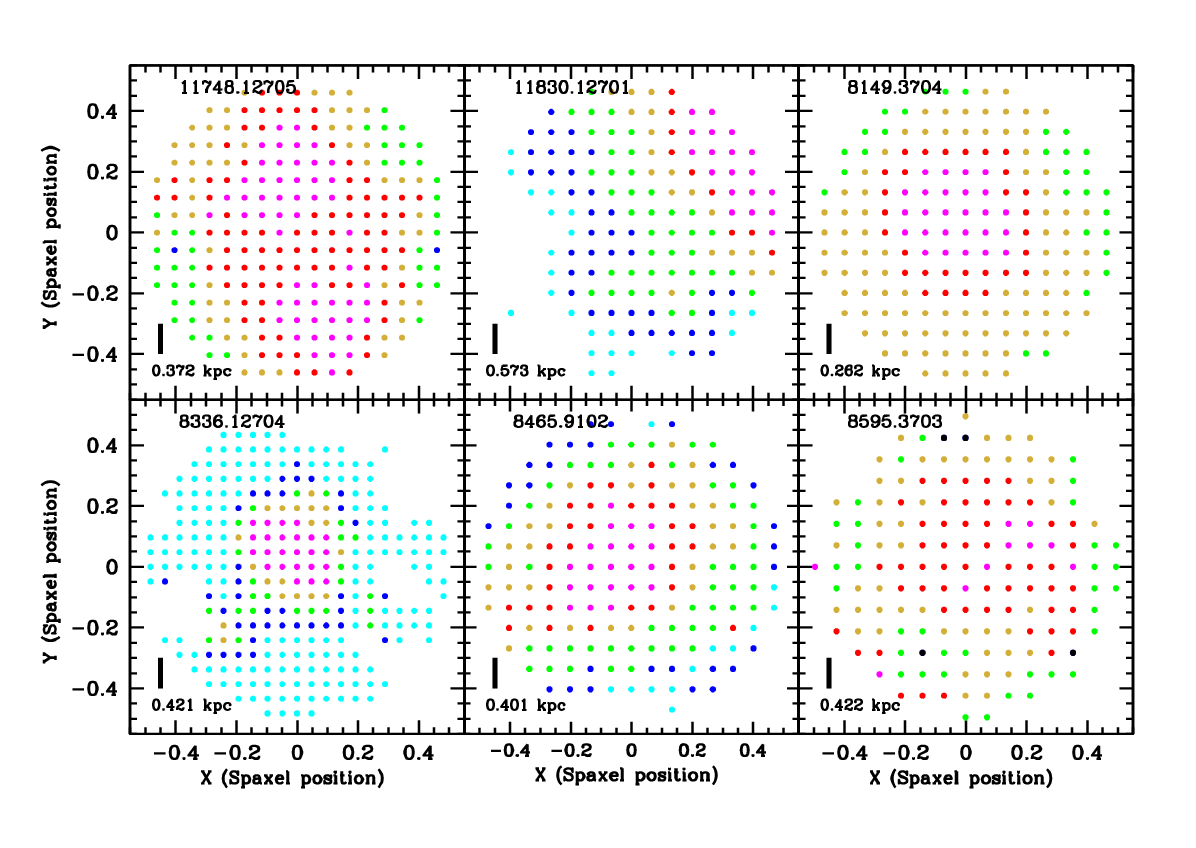}
\caption{As in the previous figure, except for the 6 galaxies with no Wolf Rayet 
detections. 
\label{models}}
\end{figure*}

\begin{figure*}
\includegraphics[width=121mm]{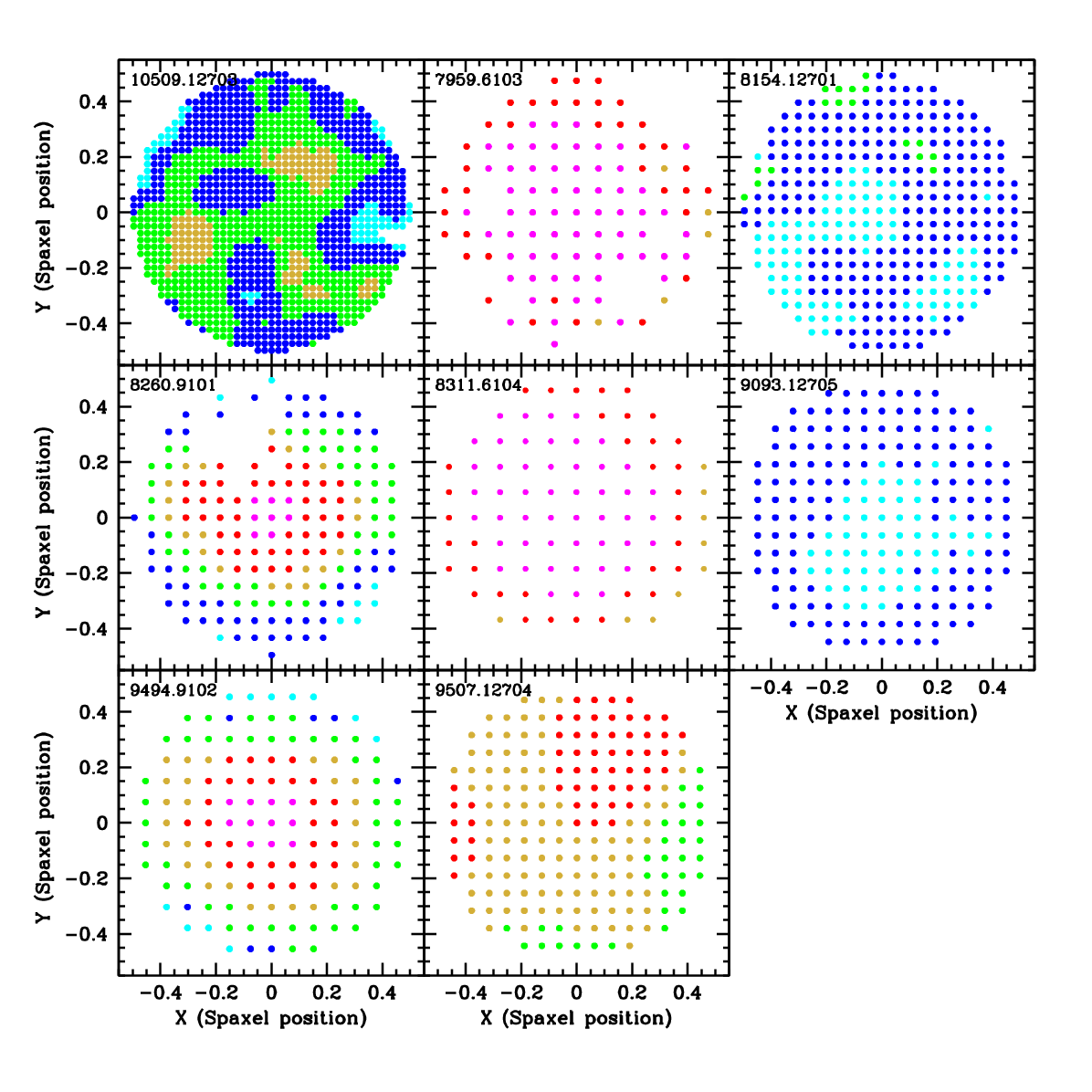}
\caption{Two-dimensional maps of [OIII]/H$\beta$  measurements for the 8 galaxies
with Wolf-Rayet detections. The circular points indicate each spaxel measurement and are
colour-coded as follows: magenta $\log$([OIII]/H$\beta$$>$0.6), red (0.4$<\log$[OIII]/H$\beta$)$<$0.6),  
dark gold (0.1$<\log$[OIII]/H$\beta$)$<$0.4), green (-0.3$<\log$[OIII]/H$\beta$)$<$0.1),
blue  (-0.5$<\log$[OIII]/H$\beta$)$<$-0.3), cyan (-0.8$<\log$[OIII]/H$\beta$)$<$-0.5).
\label{models}}
\end{figure*}
\begin{figure*}
\includegraphics[width=121mm]{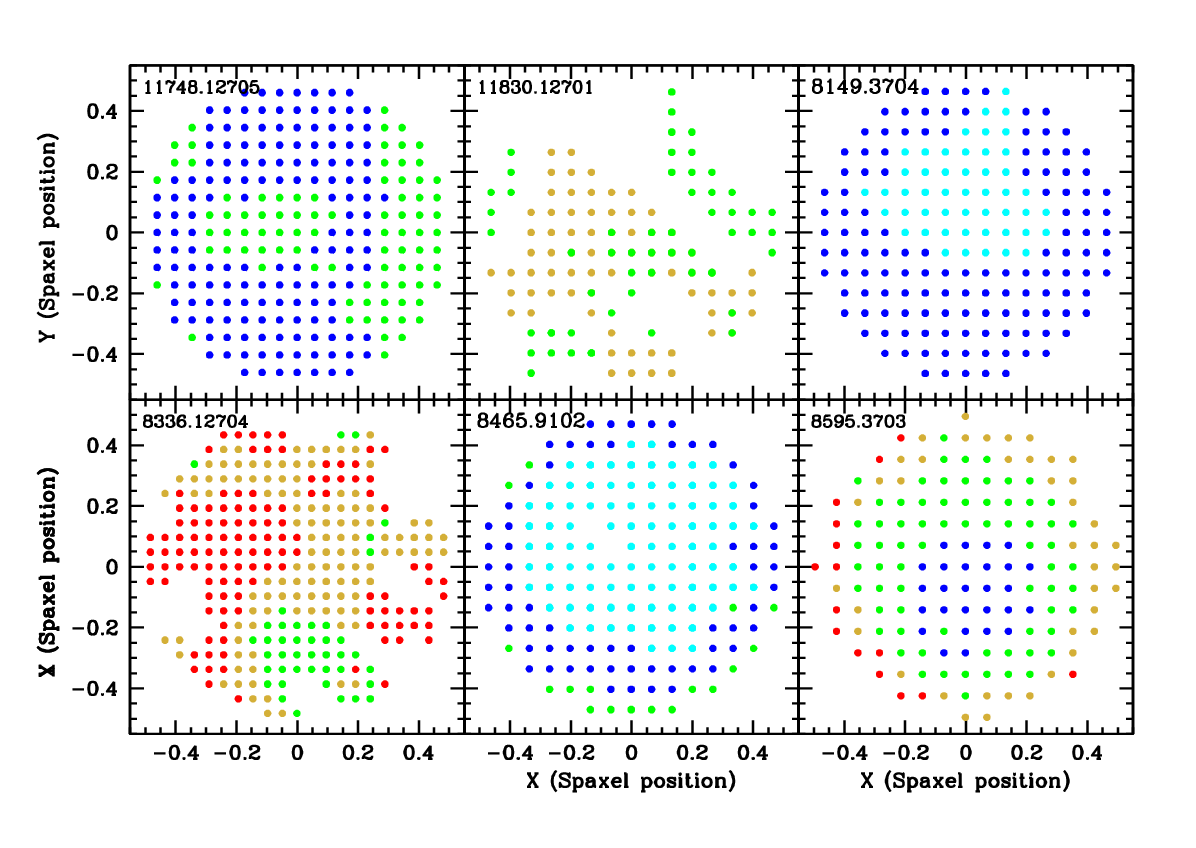}
\caption{As in the previous figure, except for the 6 galaxies with no Wolf Rayet 
detections. 
\label{models}}
\end{figure*}

\section {Wolf-Rayet Bump Regions}
                               
In this section, we present results from the stacked spectra, with emphasis on the Wolf-Rayet features
and high ionization emission lines that may indicate the presence of accreting black holes. 
In all the plots, coloured lines indicate stacked spectra of H$\alpha$ excess spaxels
selected in the plane of H$\alpha$ versus S$_{32}$ (spectra plotted in red) or 
H$\alpha$ versus [OIII]/H$\beta$ (spectra plotted in magenta). The black lines indicate 
stacked spectra for the spaxels selected to have normal values of EQW H$\alpha$. 

Figure 9 shows
the blue-bump region of the spectra. Clear broadened excess HeII$\lambda$4686 emission is
visible in  5 galaxies (7959.6103, 8260.9101, 8311.6104, 9093.12705, 9507.12704).
The object 7959.6103 shows the strongest ($>$ 10\%) excess. Objects 8260.9101, 8311.6104, 
9093.12705  and 9507.12704 show smaller broad line  excesses in the range 3-5\%. 
Strong narrow line excess HeII$\lambda$4686 emission is visible in one  
object 9494.9102. Galaxies   10509.12703 and 8154.12701 exhibit little or no    
excess emission  around the  HeII$\lambda$4686 line, but both objects exhibit very clear excesses in the
red bump region of the spectrum, so they are included in our list of galaxies with 
Wolf-Rayet feature detections.  

Figure 10 shows
the red-bump region of the spectra. CIV$\lambda$5801 is the clearest
Wolf-Rayet emission line feature in this wavelength range and is clearly in excess in
objects 10509.12703, 8154.12701, 8260.9101, 8311.6104, 9494.9102 and 9507.12704. 
Excess CIII$\lambda$5696 is also sometimes seen, but less frequently than CIV$\lambda$5801.
Unlike the blue-bump excess, which is confined almost entirely to
a narrow wavelenth interval  around the HeII$\lambda$4686 emission lines, 
the excess red bump emission extends over a wavelength interval of 
100-150 \AA\ in most galaxies and the  typical flux enhancement is at the  3-5\%
level across this region. As we will discuss in the next section, not all  the excess emission
may  be   
of stellar origin, because there are clear indications for very high ionization lines
in some  galaxies that are generally believed to be signatures of excitation
of gas around accreting black holes.

In Appendix B,  
Figures B1 and B2 show stacked spectra in the blue and red bump wavelength regions for the sub-sample listed in
Table 1 as having no excess WR feature detections. 
In these objects, the emission close to the HeII$\lambda$4686  or the CIV$\lambda$5801 lines is the same in
the stacked H$\alpha$ excess spectrum and in the stacked spectrum of the normal star-forming regions 
to within  2$\sigma$.

\begin{figure*}
\includegraphics[width=105mm,angle=270]{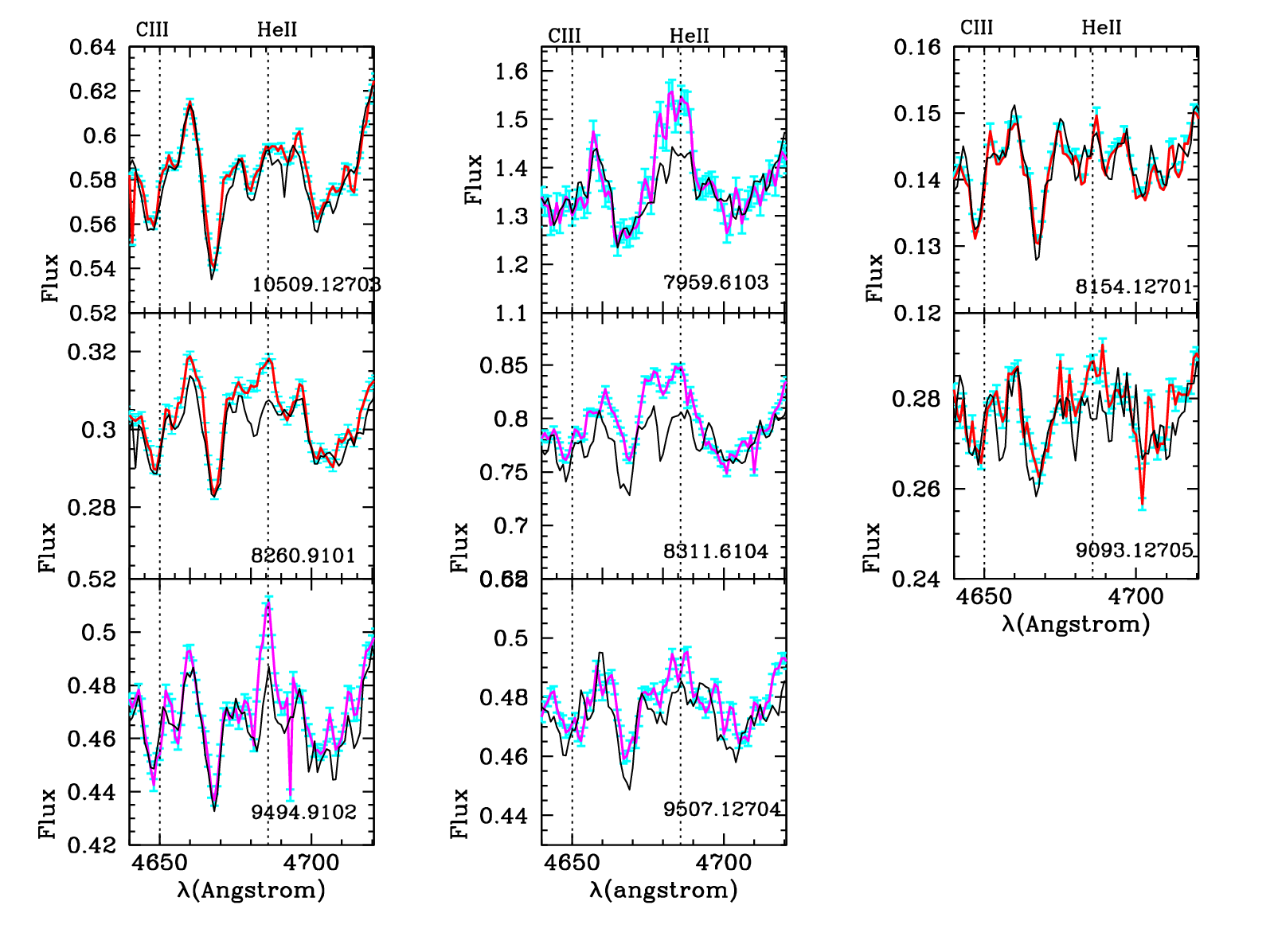}
\caption{The ``blue bump'' region of the stacked spectra is shown  for the selected H$\alpha$ excess
spaxels (red or magenta lines) compared to the spaxels with normal starburst values of the   H$\alpha$ equivalent
(black lines). Errorbars on the  H$\alpha$ excess spectra are plotted in cyan.      
The spectra plotted in red are selected using a cut on S32, while the spectra plotted in magenta are
selected using a cut on $\log$ [OIII]/H$\beta$. The dotted lines show the location of the emission
lines HeII and CIII as  labelled on the top panels. 
\label{models}}
\end{figure*}
\begin{figure*}
\includegraphics[width=105mm,angle=270]{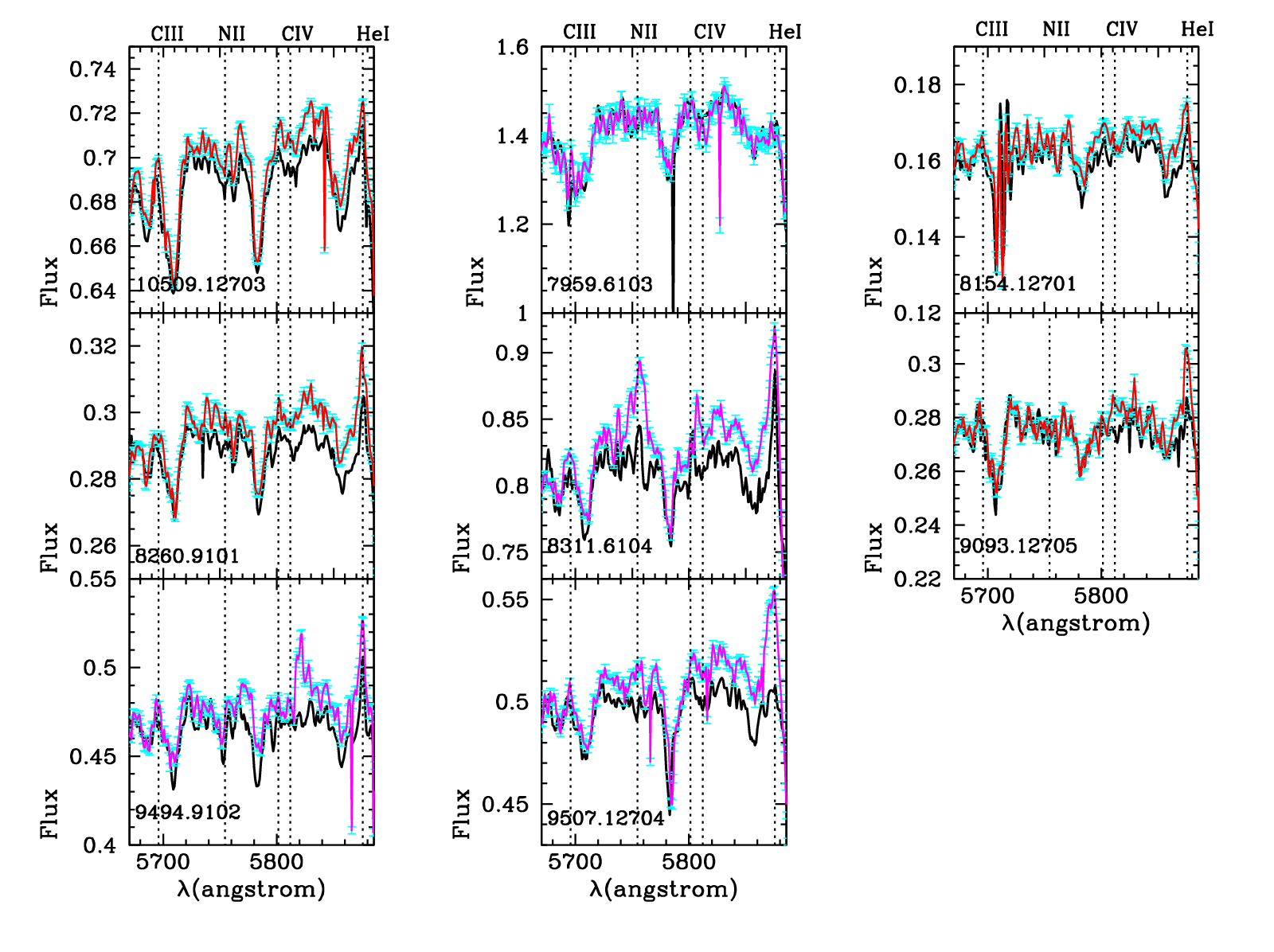}
\caption{As in the previous figure, except for the ``red bump'' region of the stacked spectra.                                                    
\label{models}}
\end{figure*}

\section {Higher-ionization emission lines indicative of accreting black holes}

Figures 11 and 12 show the regions of the spectrum covering some of the Balmer
absorption line features and the high-excitation [NeIII]$\lambda$3869 emission line.
for the 8 galaxies with Wolf-Rayet feature detections (Fig. 11) 
and the 6 galaxies without detections (Fig. 12). Neon is produced during the late stages of massive stellar evolution, 
by carbon burning and is expected to closely track oxygen abundance. Compared to doubly ionized Oxygen, which
has an ionization potention of  35.12 eV, doubly ionized Neon with its ionization potential of 40.96 eV 
will be more sensitive to the presence of higher energy ionizing photons. This sensitivity is
the reason why this line is sometimes  used in          
diagnostics distinguishing star-forming galaxies from galaxies with active galactic 
nucleus (AGN) activity (Rola et al. 1997; Perez-Montero et al. 2007). 

In the galaxies without WR detections,  the Balmer absorption features   
differ very little between the stacked spectra covering the H$\alpha$ excess and the normal starburst regions.
In two  objects (8149.3704 and 8465.9102 ), the Balmer absorption lines are weaker and there
is clear emission present in the higher order H$\epsilon$ and  H8 Balmer absorption features in the
spectra covering the  H$\alpha$ regions, showing that star formation activity is present.  
In the other objects, there are no clear differences in the Balmer absorption
lines, most likely because the young ionizing stars are too heavily obscured by dust to be seen in
the spectra. The  [NeIII] lines are weak in all 6 galaxies without WR detections
and there is no difference between the line strengths in the  H$\alpha$ excess and
 normal starburst stacked spectra.

In contrast, Figure 11 shows  that the stacked spectra
of the  H$\alpha$ excess spaxels in the sample with Wolf-Rayet detections exhibit enhanced  [NeIII]
emission in five out of 8 galaxies (10509.12703, 8260.9101, 8311.6104, 9494.9102 amd 9507.12704). In these galaxies,
the Balmer absorption lines are often weaker, without clear indication of infilling by  a narrow emission line component.
If the origin of the [NeIII] excess is from AGN, it could be that the weaker Balmer emission is
produced by a non-thermal component (e.g. radiation from an accretion disk).  

We further investigate evidence for accreting black holes 
by looking for [FeX]$\lambda$6374  coronal line emission. This transition has an ionization
potential of 262.1 eV and is therefore considered  a  reliable signature of AGN activity in galaxies 
(e.g., Penston et al. 1984; Prieto \& Viegas 2000). Coronal line emission can be produced either by gas 
photoionized by a  hard AGN continuum (e.g., Korista \& Ferland 1989; Oliva et al. 1994; Pier \& Voit 1995)
or mechanically excited by radiative shock waves that are driven into the host galaxy by 
radio jets from the AGN (Wilson \& Raymond 1999). Recently, Molina et al (2021) and Kimbro et al (2021) have used 
this line as a way of searching for black holes in dwarf galaxies.  
In Figure 13, we present stacked spectra from 4 galaxies where there is  a clear feature at wavelengths
close to 6374 \AA. The stacked spectra from the H$\alpha$ excess spaxels are plotted in red and
the stacked spectra from the normal starburst spaxels are plotted in black. We note that we also searched
the spectra of the 6 galaxies with no Wolf-Rayet  detections for emission lines in the neighbourhood of  6374 \AA
emission and did not find any clear detections.   

One possible contaminant is  [OI]$\lambda$6363, which  is expected to be about 3 times weaker than
[OI]$\lambda$6300 for starburst galaxies (Heckman, private communication). In order to distinguish
between [OI] and [FeX], we plot the spectra over the wavelength range that includes [OI]$\lambda$6300.
The emission features in the two galaxies   8311.6104 and 9507.12704 are quite consistent with [OI] emission.

The galaxy 10509.12703 exhibits a weak [FeX]$\lambda$6374 line  centered at 6374 \AA\ rather than 6363 \AA\,
as is the case in the galaxies in the two bottom panels.  
There is also a very strong absorption feature  bluewards of the
emission line that is seen in both the nuclear and in the outer regions of the galaxy. 
It is possible that the [FeX]$\lambda$6374 emission line  may trace shocked outflowing gas in this system.
The galaxy is not interacting and  the ionized gas kinematics is dominated by rotation. We note that the   
[SIII]$\lambda$9531 line is narrow and symmetric even in the  H$\alpha$ excess regions in this object, so
if there is an outflow, it is confined to the extremely highly excited gas.

The galaxy 7959.6103 exibits weak but broad emission that spans 
the wavelength range between   [OI]$\lambda$6363,and  [FeX]$\lambda$6374 emission. Interestingly, the
emission line is narrower in the stacked spectrum covering the
H$\alpha$ excess region (red spectrum) 
and it is also more clearly centred at 6374 \AA. 
This galaxy is rather different from most of the others
studied in this paper in that the H$\alpha$ excess spaxels are  fewer in number
and located entirely outside the central region of the galaxy. Their stacked spectrum exhibits a very
strong broad HeII$\lambda$4686 line excess, but there is  no red bump or   [NeIII]$\lambda$3869
emission line excess.  Although the galaxy has a companion, the  H$\alpha$ gas kinematics is
regular and rotation dominated. The galaxy also has an earlier type morphology and
all the emission lines are significantly weaker in this system than in most of the others.       
This suggests that this  galaxy is a system where there are
isolated pockets of Wolf Rayet stars where gas has been locally compressed and shocked.
As we will discuss in section 8, 
if the starburst extends across a large region
within the galaxy, it is likely that  blue and red bump features will be seen simultaneously. If the
starburst is confined to a very small region, this may not be the case because the
blue and red bump features do not peak contemporaneously.

We note that the radio emission in the two objects with tentative [FeX] line detections is
weak (see Figure 1) and they are also not the systems where the stacked H$\alpha$ excess spaxel spectra
exhibit  strong excess [NeIII]$\lambda$3869 emission , suggesting that these two lines may be excited
by different physical mechanisms.

\begin{figure*}
\includegraphics[width=117mm,angle=270]{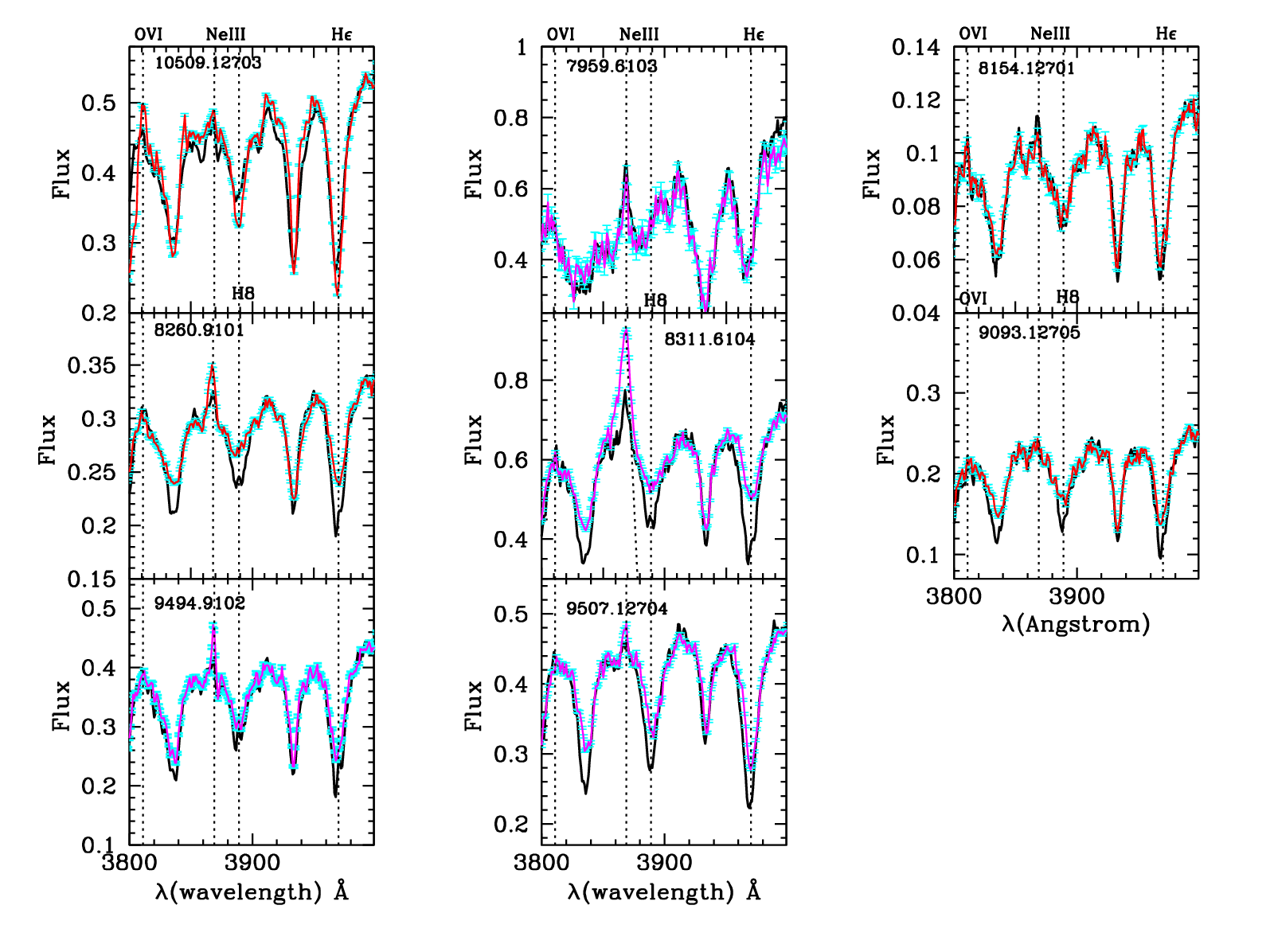}
\caption{The Balmer absorption line  region of  stacked spectra from $\alpha$ excess
spaxels (red or magenta lines) is  compared to stacked spectra from spaxels with normal starburst values of the  H$\alpha$ equivalent
width (black lines). Results are shown for the 8 galaxies with WR feature detections. 
Errorbars on the  H$\alpha$ excess spectra are plotted in cyan.
The spectra plotted in red are selected using a cut on S32, while the spectra plotted in magenta are
selected using a cut on $\log$ [OIII]/H$\beta$.
\label{models}}
\end{figure*}

\begin{figure*}
\includegraphics[width=85mm,angle=270]{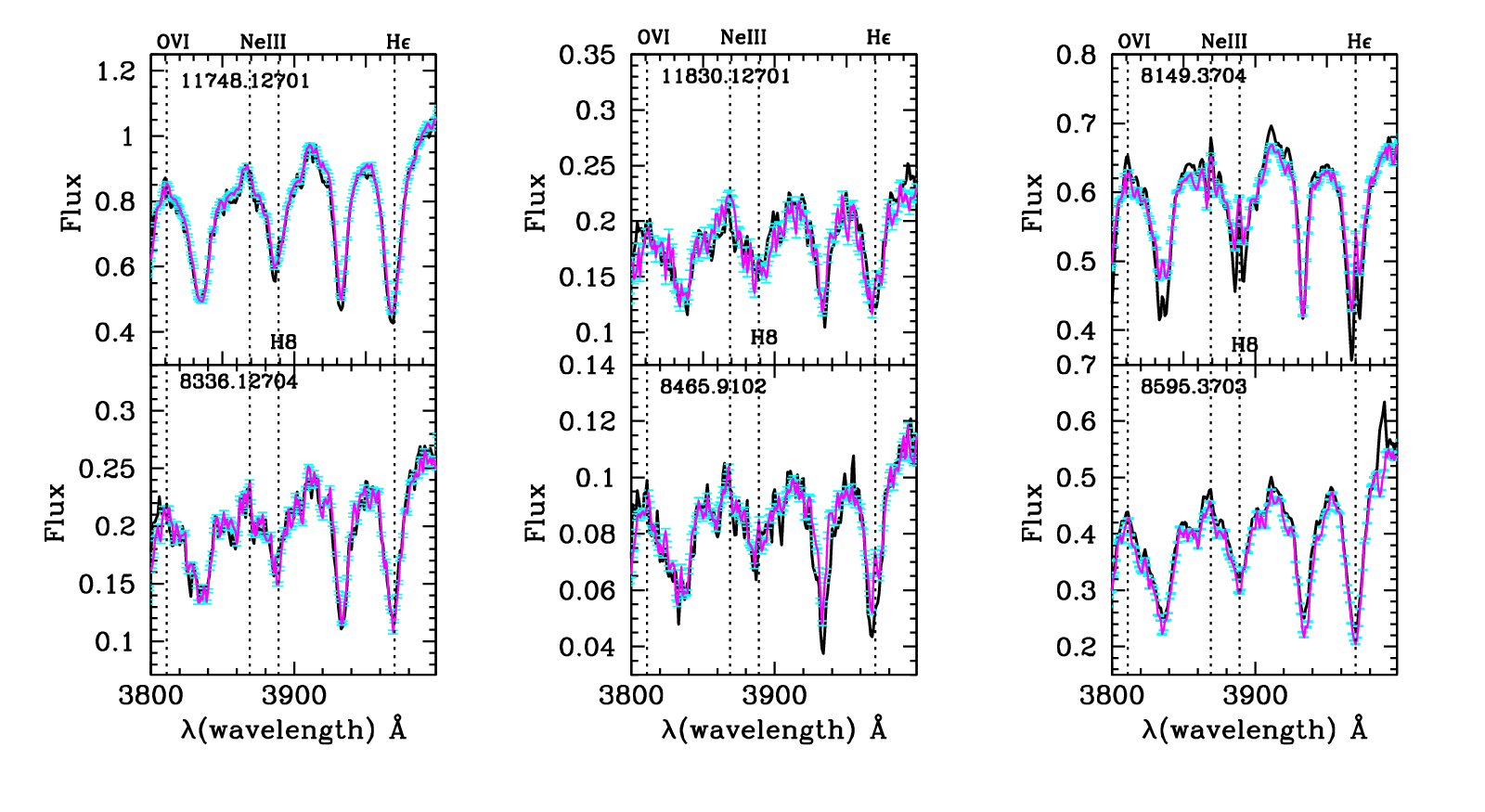}
\caption{As in the previous figure, except comparing stacked spectra from  H$\alpha$ excess spaxels in the 6 galaxies with no WR feature detections 
with  stacked spectra from normal starburst spaxels. Results are shown for 
the 8 galaxies with WR feature detections.
Errorbars on the  H$\alpha$ excess spectra are plotted in cyan.
\label{models}}
\end{figure*}

\begin{figure}
\includegraphics[width=84mm]{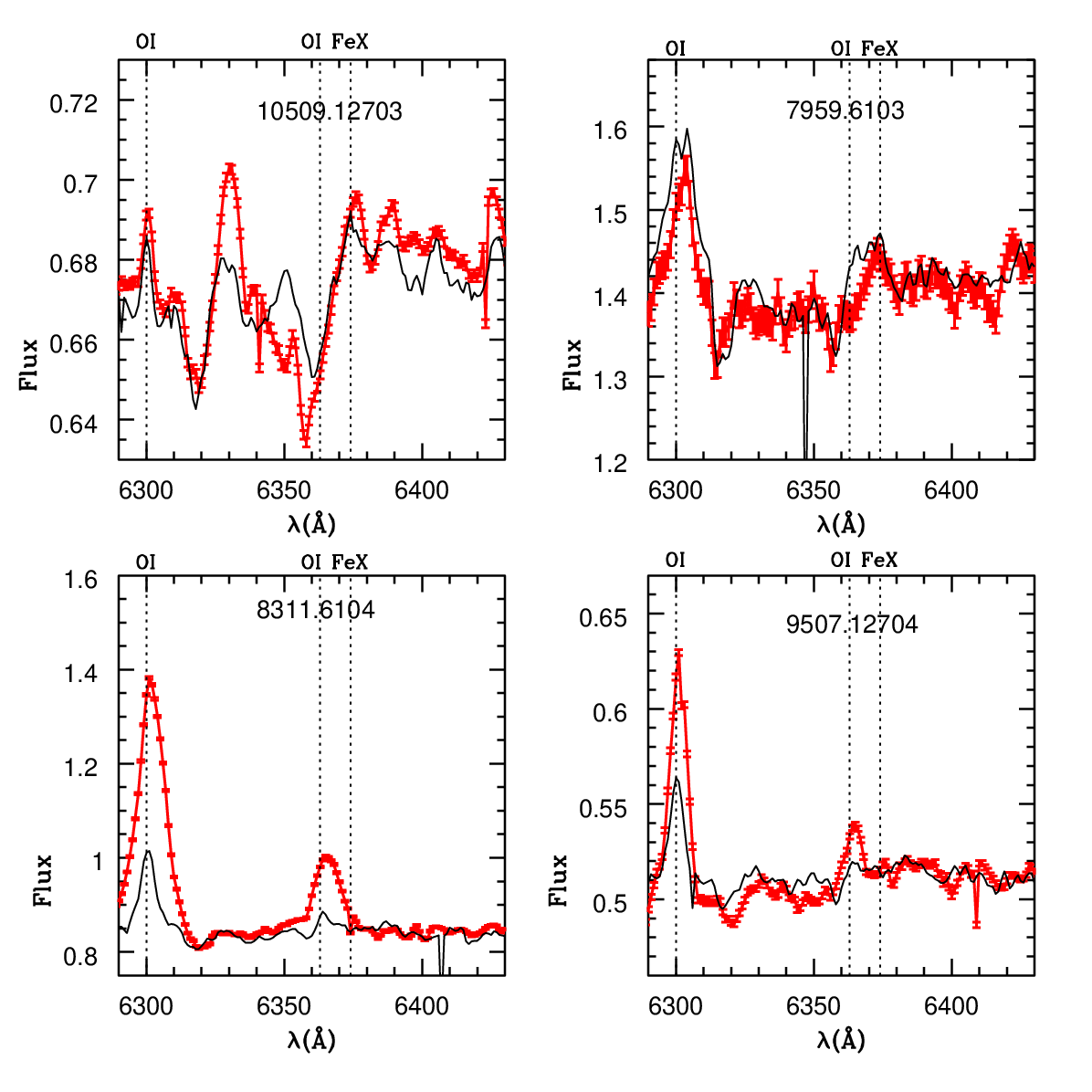}
\caption{The region of the spectrum containing the [FeX] coronal line.
This line is often used as an AGN indicator, though it can also be produced
in SNII shocks. Four galaxies from the sample with WR detections have [FeX] detections
and spectra in the regions of this
line are  plotted in the figure. The stacked spectrum from the H$\alpha$ excess spaxels
is plotted in red and the stacked spectrum from the normal starburst spaxels
is plotted in black.  The sample without WR detections was checked and no [FeX]
line detections were found. The location of the two [OI] lines at 6300 and 6363 \AA are also
indicated on each panel.
\label{models}}
\end{figure}

\section {Local ISM properties}

So far we have not found any clear explanation for why only some galaxies in the H$\alpha$-selected
sample show signatures of Wolf-Rayet stars. We have shown that Wolf-Rayet star regions are
often associated with regions of the galaxy  with very high-ionization line emission, but
this is most likely a  consequence of the the presence of these young massive stars
(and possible associated accreting black holes),
not an explanation of what causes them to form in the first place. We also showed
that the ionized gas kinematics traced by the H$\alpha$ emission line within a radius $R_{50}$
was indistinguishable  in the 
two  sub-samples of galaxies with and without Wolf-Rayet detections (see Figures A3
and A4).  

We now turn to an examination of colder neutral gas in these galaxies. We make use of the 
Na I $\lambda\lambda$5890,5896 absorption line -- the ionization potential of NaI is only 
5.1 eV, so this absorption feature probes the neutral HI phase of the interstellar
medium. It was first used as a probe of the neutral gas in a
large sample of  starburst galaxies by Heckman et al (2000). 
The galaxies were selected on the basis of far-IR flux and very warm far-IR color temperatures. 
The longslit spectra typically covered the inner few hundred to few thousand parsecs
of each galaxy.
In most  galaxies, the absorption feature
had a smaller velocity width than the stellar component in the same region,
so dynamically cold interstellar gas 
was interpreted to be the main  contribution to the observed NaD line.
The NaD line was also  found to be blueshifted by $\Delta v >  100$ km/s relative 
to the galaxy systemic velocity in 12 out of the 18 cases, indicating that there
were cold gas outflows in these galaxies. Finally, there was a strong correlation
between the reddening of the stellar continuum and the depth of the absorption.
Heckman et al interpreted these observations in terms of an absorbing screen of gas
and dust produced by gas outflowing 
along the minor axis of the the galaxy.

In this work, we  compare the NaD absorption feature in the stacked spectrum
covering the H$\alpha$ excess region of the starburst with that in the
stacked spectrum covering the normal starburst regions. The comparison is
thus of {\em local} ISM conditions in different regions of the same galaxy.  
In addition we compare results for the sub-samples with and without
Wolf-Rayet feature detections. Figure 14 shows the stacked spectra in the region of the
NaD doublet for the 8 galaxies with Wolf-Rayet feature detections.  
As can be seen, the doublets in the  H$\alpha$ excess regions are almost
always stronger than in the normal starburst regions. The only exception is galaxy 
7959.6103. 

We have measured the equivalent width of the NaD feature and the results
are summarized in Figure 15.
In the top left panel, the magenta triangles show the EQW 
measurements for the NaD absorption doublet
in the stacked spectrum covering the  H$\alpha$ excess spaxels for each of the 8 galaxies
with WR feature detections. The black pinwheel  symbols 
show the correspnding NaD EQW for the stacked
spectrum of the normal starburst spaxels in each object. The difference between
the magenta triangle and the black pinwheel at the same location of the x-axis is
a measure of the change in NaD EQW between the two regions, which is typically
around 2 \AA\ in this subsample. In the top right panel, we show the same plot
for the subsample without Wolf-Rayet feature detections. The shift 
is considerably smaller, typically around 0.5 \AA. 
Since the total range in NaD EQW is very similar in the two subsamples,
we interpret these results as saying that greater cold gas {\em compression} 
may be important in explaining the presence of excess Wolf-Rayet stars.  

In the bottom left panel, we examine whether there is a  correlation between
NaD EQW and dust in the H$\alpha$ excess regions of starburst galaxies
Red triangles are for the 8 galaxies with WR detections and
 black triangles are for the 6 galaxies without WR detections.  
No clear correlation is found in either sample. As discussed earlier,
this is in strong contrast to the IR-selected starburst sample
studied by Heckman et al 2000. 
These results indicate that the dust and cold gas do not
form a foreground screen in H$\alpha$-selected starbursts, but are more well mixed
with the stars. We also note the the NaD doublets in the H$\alpha$ excess
galaxies  do not exhbit significant
blue-shifts (see Figure 14) indicative of outflows.

In the bottom right panel, the ratio of the NaD  
central flux decrement (defined as the flux at the centre of the line divided by the
continuum flux) is plotted as a function of NaD equivalent width for the same
two samples. Only a weak
correlation is seen and the flux decrements  do not extend to the very low
values ($\sim$0.1-0.2) found in the sample studied in the Heckman et al (2000) paper,
again indicating a deviation of ISM conditions in these galaxies from
IR-selected  starbursts.

\begin{figure}
\includegraphics[width=71mm,angle=270]{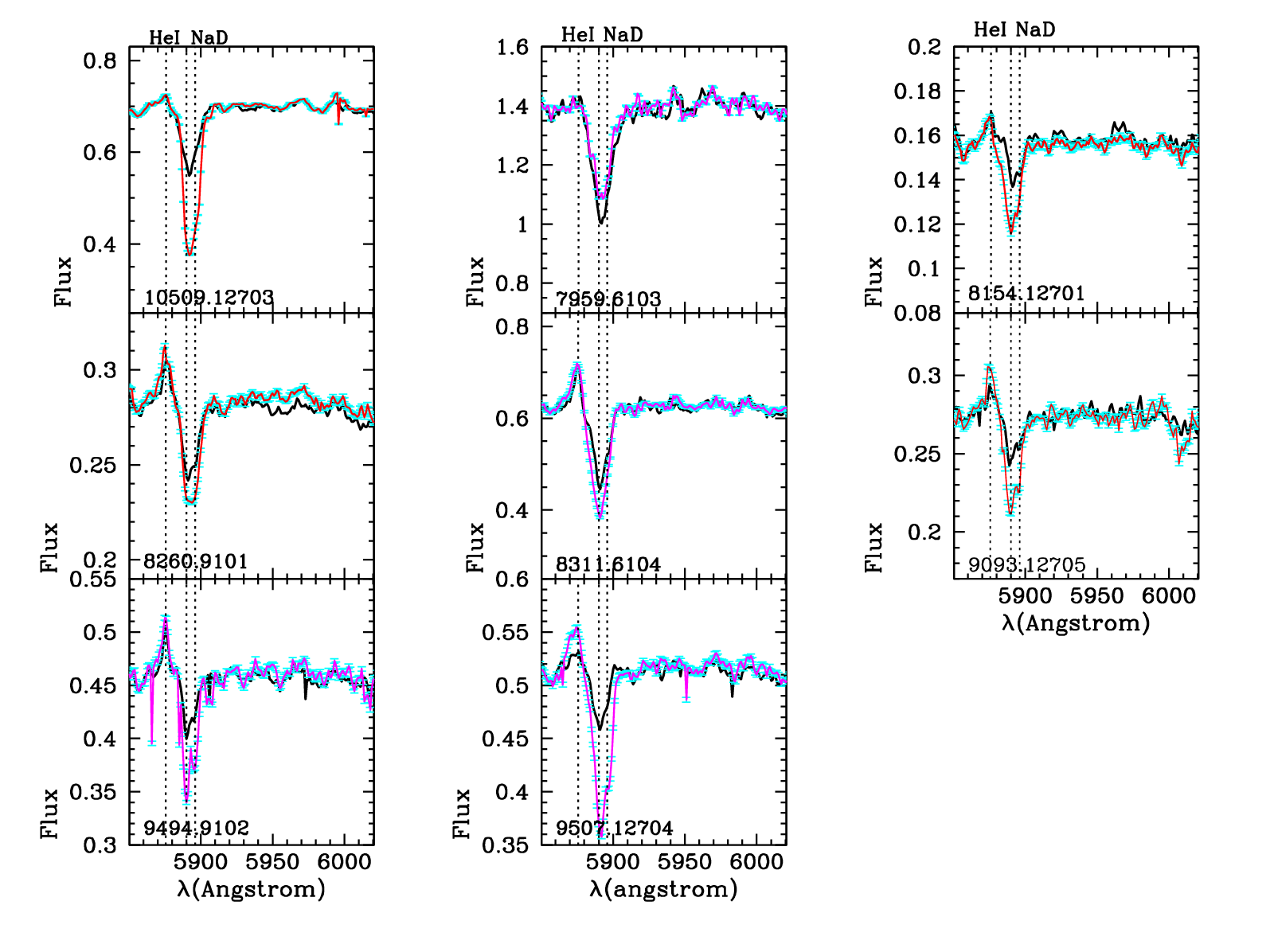}
\caption{The region of the stacked spectra from  H$\alpha$ excess
spaxels containing the NaD absorption line doublet 
(red or magenta lines) is  compared to stacked spectra for  spaxels with normal starburst values of the  H$\alpha$ equivalent
width (black lines). Results are shown for the 8 galaxies with WR feature detections. 
Errorbars on the  H$\alpha$ excess spectra are plotted in cyan.
\label{models}}
\end{figure}

\begin{figure}
\includegraphics[width=86mm,angle=0]{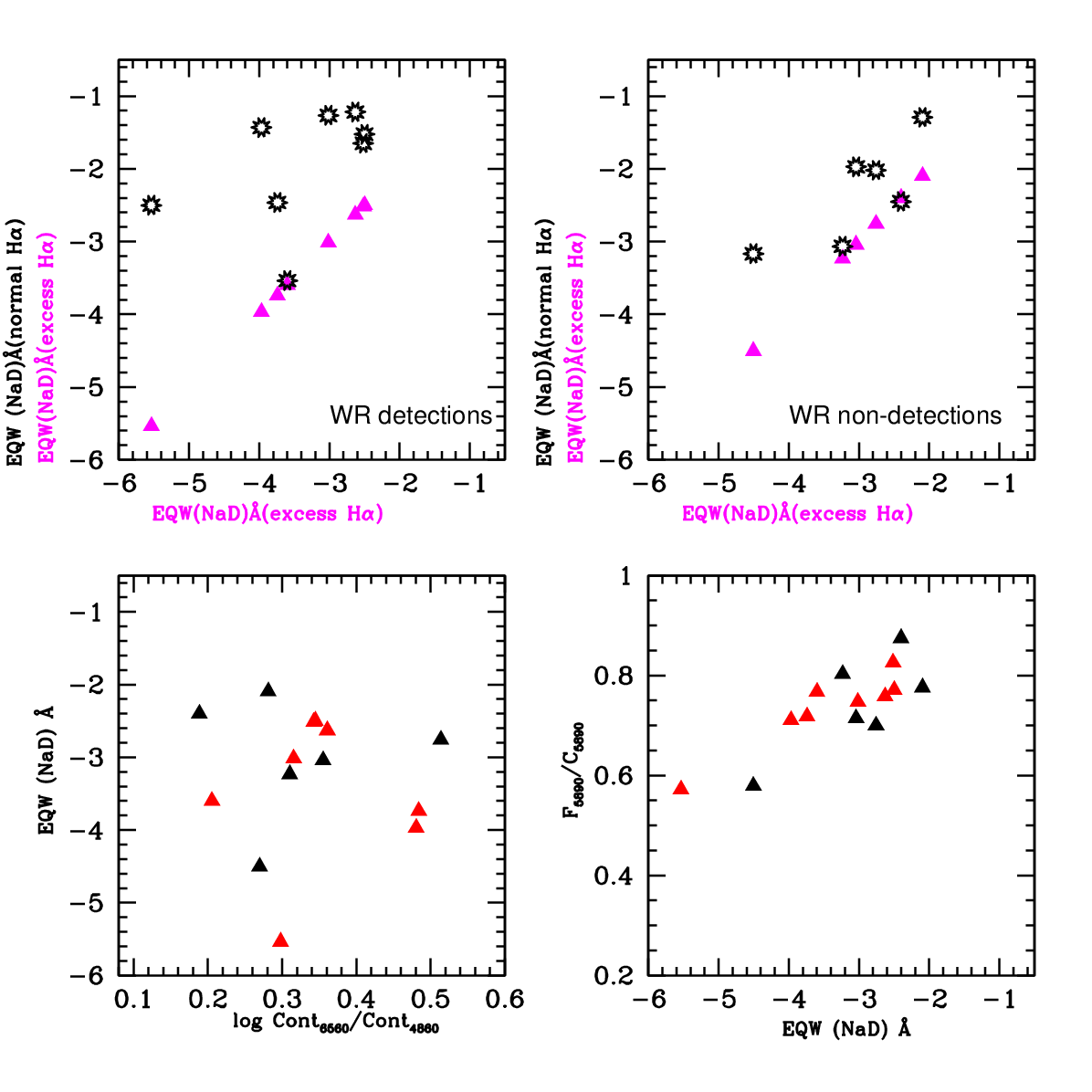}
\caption{{\bf Top panel left:} The magenta triangles show the EQW measurements for the NaD absorption doublet
in the stacked spectrum for the  H$\alpha$ excess spaxels for each of the 8 galaxies
with WR feature detections. The black symbols show the correspnding NaD EQW for the stacked
spectrum of the normal starburst spaxels in each object. {\bf Top panel right :} Same as the top
left panel, except for the 6 galaxies without WR detections.
{\bf Bottom left:} The NaD EQW in H$\alpha$ excess regions  is plotted against a measurement of the reddening in the
stellar continuum. 
Red triangles are for the 8 galaxies with WR detections
while black triangles for for the 6 galaxies without WR detections.  
{\bf Bottom right:} the ratio of the NaD
central flux decrement (defined as the flux at the centre of the line divided by the
continuum flux) is plotted as a function of NaD equivalent width for the same
two samples. 
\label{models}}
\end{figure}

To gain further insight, we examine local ISM conditions in the same way using the
[SIII]$\lambda$9531 emission line previously 
used for selecting regions
likely to contain Wolf-Rayet stellar features.
Figure 16 shows the stacked spectra in the region of the
line for the 8 galaxies with Wolf-Rayet feature detections.
The line is always stronger in the H$\alpha$ excess stack except, once again, for galaxy 7959.6193.
The line also varies in width and exhibits  a blue-side asymmetry in some cases.

Results on the properties of the [SIII]$\lambda$9531 line are shown quantitatively 
in Figure 17. In the top panel, the full width half
maximum (FWHM) of the line is plotted against the line equivalent width.
Results are only shown for the stacked spectra covering
the  H$\alpha$ excess regions of the galaxies. Results for the sub-sample with detected Wolf-Rayet
features are plotted as magenta triangles and results for the sample without Wolf-Rayet detections
are plotted as black triangles. As can be seen, the FWHM for  the galaxies in the
sample without Wolf-Rayet detections is around 200 km/s. Half of the galaxies in the
sample with Wolf-Rayet detections have larger FWHM line widths. The largest line-widths
are found for the lowest equivalent width systems. In the top right panel, the ratio of the flux
bluewards of line centre to the flux redwards of line centre is plotted as a function of
line equivalent width. This ratio is greater than 1 for all the galaxies and reaches values
of around 2 or greater for 5 out of 8 systems with Wolf-Rayet features and 2 out of 6 galaxies
without WR features \footnote{ Note, however, that the 2 galaxies without WR features 
with large values of this asymmetry
paramter have small equivalent width, and the measurements may be more affected by continuum subtraction
errors}.  

In the bottom left panel of Figure 17, the [SIII]$\lambda$9531 
equivalent width is plotted as a function
of the same continuum reddening parameter as in the bottom left panel of Figure 15. In this case,
there is a clear correlation with continuum reddening. The H$\alpha$ excess galaxies
tend to have higher [SIII] equivalent widths and more reddening. 
In the bottom right panel of Figure 17, we plot the
FWHM of the line as a function of  continuum reddening. We find an anti-correlation in the
sense that the less reddened systems have higher line widths.  These results may indicate that the
the dust is being expelled together with the ionized gas, and that the neutral gas remains locked
in the interstellar medium within the disk. 

In summary, we conclude that the H$\alpha$ excess regions of the galaxy are associated
with higher neutral gas densities than normal starburst regions. If Wolf-Rayet stars are present, this difference is even
larger.
We interpret this as saying that Wolf-Rayet stars 
are produced when the ambient interstellar gas is compressed to unusually high densities.
There is clear evidence of  outflowing, dusty ionized gas associated with  WR excess regions in the disk,
but little evidence of outflowing neutral gas.

\begin{figure}
\includegraphics[width=71mm,angle=270]{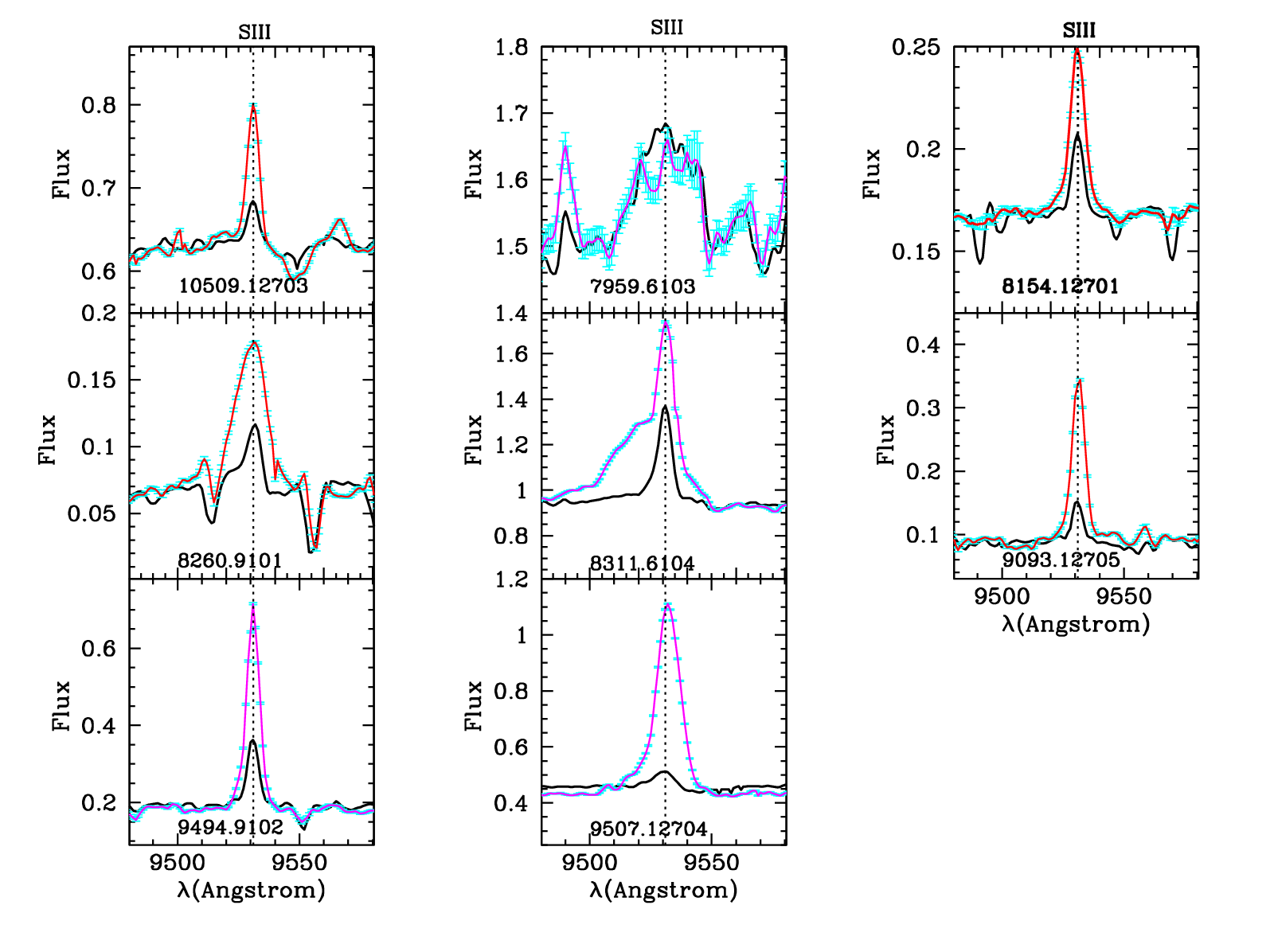}
\caption{The region of the stacked spectra from  H$\alpha$ excess
spaxels containing the [SIII]$\lambda$9531 emission line   
(red or magenta lines) is  compared to stacked spectra for  spaxels with normal starburst values of the  H$\alpha$ equivalent
width (black lines). Results are shown for the 8 galaxies with WR feature detections. 
Errorbars on the  H$\alpha$ excess spectra are plotted in cyan.
\label{models}}
\end{figure}

\begin{figure}
\includegraphics[width=86mm,angle=0]{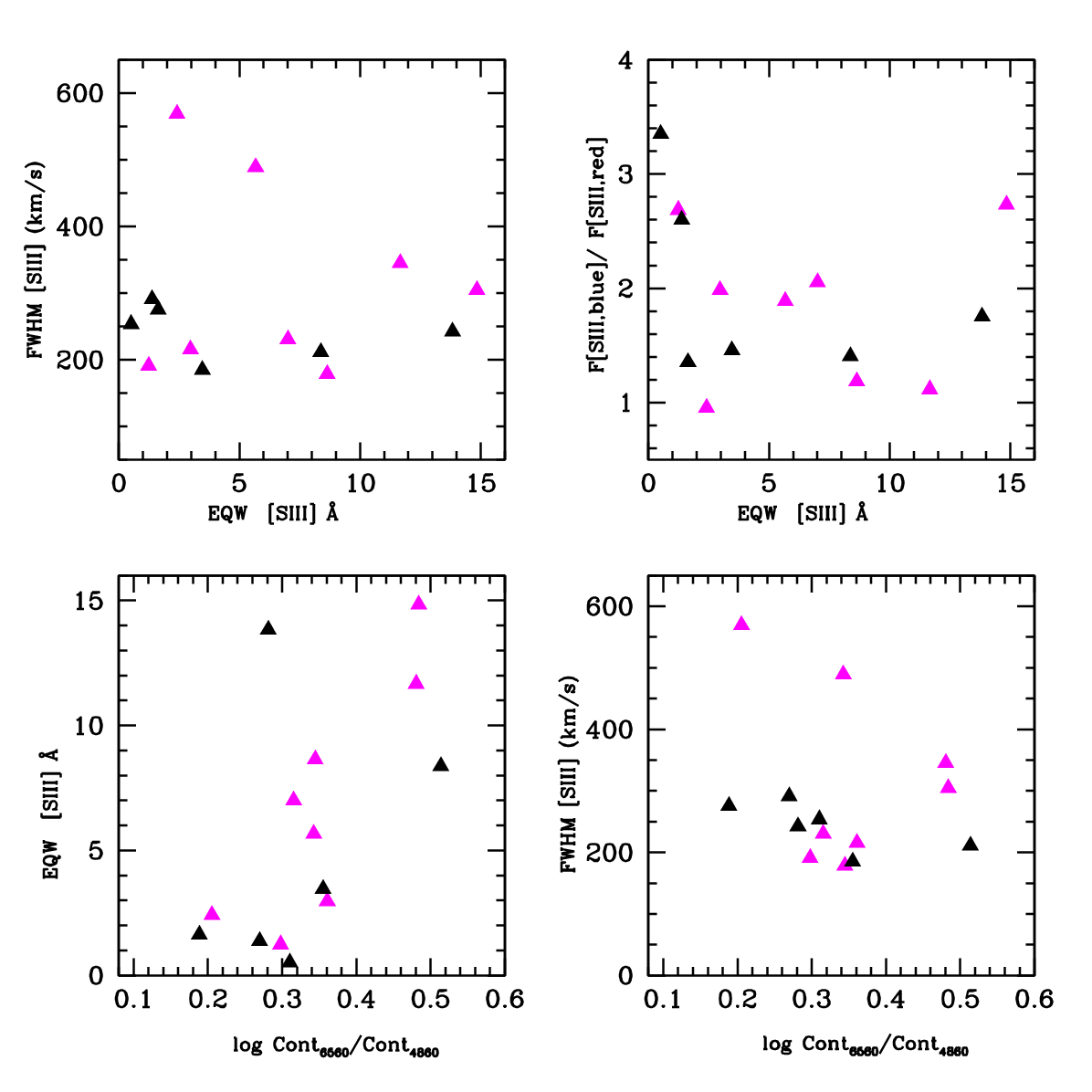}
\caption{{\bf Top left:} The full-width-half-maximum line width of the [SIII]$\lambda$9531 emission line
is plotted against the line equivalent width in the stacked spectra for the  H$\alpha$ excess spaxels.
Magenta symbols show results for the eight galaxies with WR feature detections, while
black symbols are for the 6 galaxies withour WR detections. {\bf Top right :} The line
asymmetry index, defined as the total flux bluewards of line centre dived by the total
flux redwards of line centre is plotted as a function of line equivalent width
for the two samples. {\bf Bottom left:} [SIII] line equivalent width is plotted as function of
the continuum reddening.   {\bf Bottom right:} the FWHM of the line is plotted
as a function of the continuum reddening.
\label{models}}
\end{figure}

\section {Modelling discussion: effect of IMF changes predicted by HR-Pypopstar}

The measurement of equivalent widths of weak WR features is only
possible if both the nebular emission and the contribution to the
stellar continuum from older stars can be accurately modelled and
subtracted. If the youngest stars reside in very dense, clumpy and
dusty regions of the interstellar medium, the problem of modelling the
stellar continuum becomes extremely challenging because continuum and
line radiative transfer effects have to be considered not only through
stellar atmospheres, but also through the surrounding dense  molecular
gas. As discussed in section 7, absorption lines from cold neutral gas
are clearly present in all the galaxies in our sample and the strength
of the absorption is higher for the galaxies with higher H$\alpha$
emission line equivalent widths.

In this paper, we do  model and subtract the stellar
continuum using libraries of template stars and then measure
emission line strengths, as is usually done in galaxies that are forming stars at moderate
rates. Instead, we look at the differential changes between regions where
the extinction-corrected H$\alpha$ emission is exceptionally strong and
regions where it has more normal values. The normal star-forming
regions are usually situated at distances of 3-4 kpc from the center
of the galaxy. The crossing time across these regions can be estimated
to be around $10^8$ years. One can use the Starburst99 code (Leitherer
at al 1999) to show that a  region that has formed stars at a constant
rate for $10^8$ years with a  Chabrier initial mass function will have
an H$\alpha$ equivalent width of around 200. This value is in  accord
with the normal starburst region spaxels that are selected for stacking
in our analysis.  In contrast, the H$\alpha$ excess regions, which have
H$\alpha$ equivalent widths in the range 800-2000,  span regions of 1-1.5
kpc at most and the crossing time is closer to $3\times 10^7$ years.

We now use the population synthesis model HR-pyPopStar
(Mill\'an-Irigoyen et al 2021) to generate predictions for the integrated
spectra of star-forming regions. 
The HR-pyPopStar models provides a complete set (in ages) of 
high resolution (HR) Spectral Energy Distributions of Single Stellar 
Populations. The model uses  recent high wavelength-resolution 
theoretical atmosphere libraries for main sequence (Coelho 2014), 
post-AGB/planetary nebulae (Rauch 2003),  and OB and  Wolf-Rayet stars 
(Hainich et al. 2019). The Spectral Energy Distributions are given 
for more than a hundred ages ranging from 0.1 Myr to 13.8 Gyr, at four 
different values of the metallicity (Z = 0.004, 0.008, 0.019 and
0.05). For the purposes of the analysis in this paper, it should be 
noted that the theoretical libraries of OB and  Wolf-Rayet stars have 
additional free parameters relating to the adopted stellar wind models.
For OB stars, the HR-pyPopStar models adopt moderate wind strengths, calculated with 
mass loss rate, $\log \dot{M} = -7.0$. For WR stars, mass-loss
rates are in the range $\log \dot{M}/(M_{\odot}/yr) = -4.3 - -6.1$ (Crowther 2002).

We have combined the SSPs to predict the spectra of star-forming
regions under the assumption that the star formation has been 
distributed uniformly over a time interval comparable to
the crossing time across the regions. In detail, the normal
star-forming regions are assumed to be represented by 
constant star formation for $10^8$ years with a Chabrier IMF.
We will call this the reference model henceforth.
We then test two hypotheses for the spectra of the H$\alpha$ excess
regions: 
\begin{enumerate}
\item Constant star formation for $3 \times 10^7$ years
with a Chabrier IMF.  
\item Constant star formation for $3 \times 10^7$ years  with an IMF where the
where the slope becomes flatter above 10 $M_{\odot}$.
\end {enumerate} 
For (ii) we have generated spectra for the IMF explored
in Kauffmann (2021) where the slope $\alpha$ in $dN/dM =m^{-\alpha}$
has a value 0.8 above 10$_{\odot}$, as well as a model with $\alpha=1.3$. 
We note that the recent study of Hosek et al (2019) found
a slope $\alpha=1.8$  for young stars in the Galactic Center, so
these IMF changes are fairly extreme. The
models include the contribution from nebular continuum emission, but do
not include nebular line emission.  We use a very simplified approach to
model the effect of differential extinction of the very youngest
stars due to dust in stellar birth clouds by generating models where the
very youngest stars with ages less than $3 \times 10^6$ years do not
contribute to the optical spectrum. This timescale is motivated by
recent work by Kim et al (2023) who used a combination of
JWST and ALMA data for the nearby galaxy NGC 628 to measure the timescale over which
young stars remain heavily obscured while embedded in
their parent molecular clouds.  As we will show, obscuration of
the youngest stars in the galaxies  causes the
Wolf-Rayet features to strengthen quite considerably in the predicted
integrated spectra of model 2, because the phase of the evolution
of massive stars associated with strong winds and mass loss occurs at
later ages ($10^7 - 3 \times 10^7$ years after the birth of the star).
Differences in continuum/line radiative transfer in the H$\alpha$ excess
and normal starburst ISM  regions are neglected in this analysis.

In Figure 18, we show how the predicted reference model spectra
compare with models 1 and 2. Results in the figure are
all calculated assuming solar metallicity. Figure 19 shows the same
for half solar metallicity models.  The model spectra are all normalized
to unity averaged over the wavelength range 4630-4750 \AA\ in the
left panels that  show the blue bump features, and over the wavelength range
5670-5900 \AA\ in the right panels that show the red bump features.
The reference model is plotted in black in each panel. The spectra plotted
as magenta/cyan dotted lines in Figures 18/19 do not include any dust extinction.
The spectra plotted as solid red/blue lines in the 
two figures assume that stars with ages
less than $3 \times 10^6$ years are completely obscured at optical
wavelengths.  

The top panels in Figures 18 and 19 show almost no difference 
between the reference model spectrum
and the spectrum of a region that  has formed stars for a shorter time interval 
with the same IMF. This is true independent of whether or not dust
extinction of the youngest stars is included.  
The middle and bottom panels show that significant (5-10 \%) flux enhancements
simular to those seen in the observations are only achieved 
for solar metallicty  models where the IMF slope flattens at high masses
and where the youngest stars are highly obscured.  Diffential extinction
is less important for producing clear Wolf-Rayet bumps for the models 
with half solar metallicity. The blue bump around the HeII$\lambda$4686
line is also stronger for the half solar model than for the solar model.
Part of this difference may be caused by the fact that the continuum
emission suffers less line-blanketing at lower metallicities,
making the wind features easier to see in the spectra.

The Kauffmann (2021) IMF  with $\alpha=0.8$ above 10$M_{\odot}$  
combined with a constant star formation rate of $3 \times 10^7$ years
results in a  predicted H$\alpha$ equivalent width of 1500, which is
nicely consistent with our adopted cut on EQW(H$\alpha$) used to define
our H$\alpha$ excess spaxels.

Some words of caution are now in order.
An important issue that is not yet addressed  is the fact that
the H$\alpha$ excess spaxels were not selected purely on the basis
of H$\alpha$ equivalent width, but in a two dimensional space of 
EQW(H$\alpha$) and ionization parameter.  Binary star systems are known
to produce harder ionizing spectra and an alternative explanation for
the observed spectral differences presented in this paper
could be a larger fraction of interacting massive
stars in the central dense ISM regions of our galaxies rather than
an excess population of stars  formed at higher masses. More detailed
modelling comparing the photo-ionizing effects of
interacting  binaries with that of  massive O stars,  
is required before further progress can be made (see for example
Götberg et al 2018).  

We also note that the width of the red bump feature predicted by the models
is comparable to the width of the excess emission in the observed spectra, but
the same is not true for the blue bump emission, which is considerably narrower in the
data than in the models. Assigning a fixed set of wind parameters
to all Wolf-Rayet stars in the input stellar library is no doubt a serious
oversimplification. Some of the WR lines such as CIII$\lambda$5696 are very
sensitive to the wind density (Crowther 2002). Finally, 
we note that the HR-pyPopStar models employ Padova isochrones
(Bressan et al. 1993;
Fagotto, et al. 1994; Girardi et al. 1996), which 
do not include the effects of stellar rotation on
the evolution of stars -- the lifetimes of massive stars are longer in these models
and this will affect the predicted spectra. More detailed examination of
these issues is deferred to future work.

\begin{figure*}
\includegraphics[width=101mm]{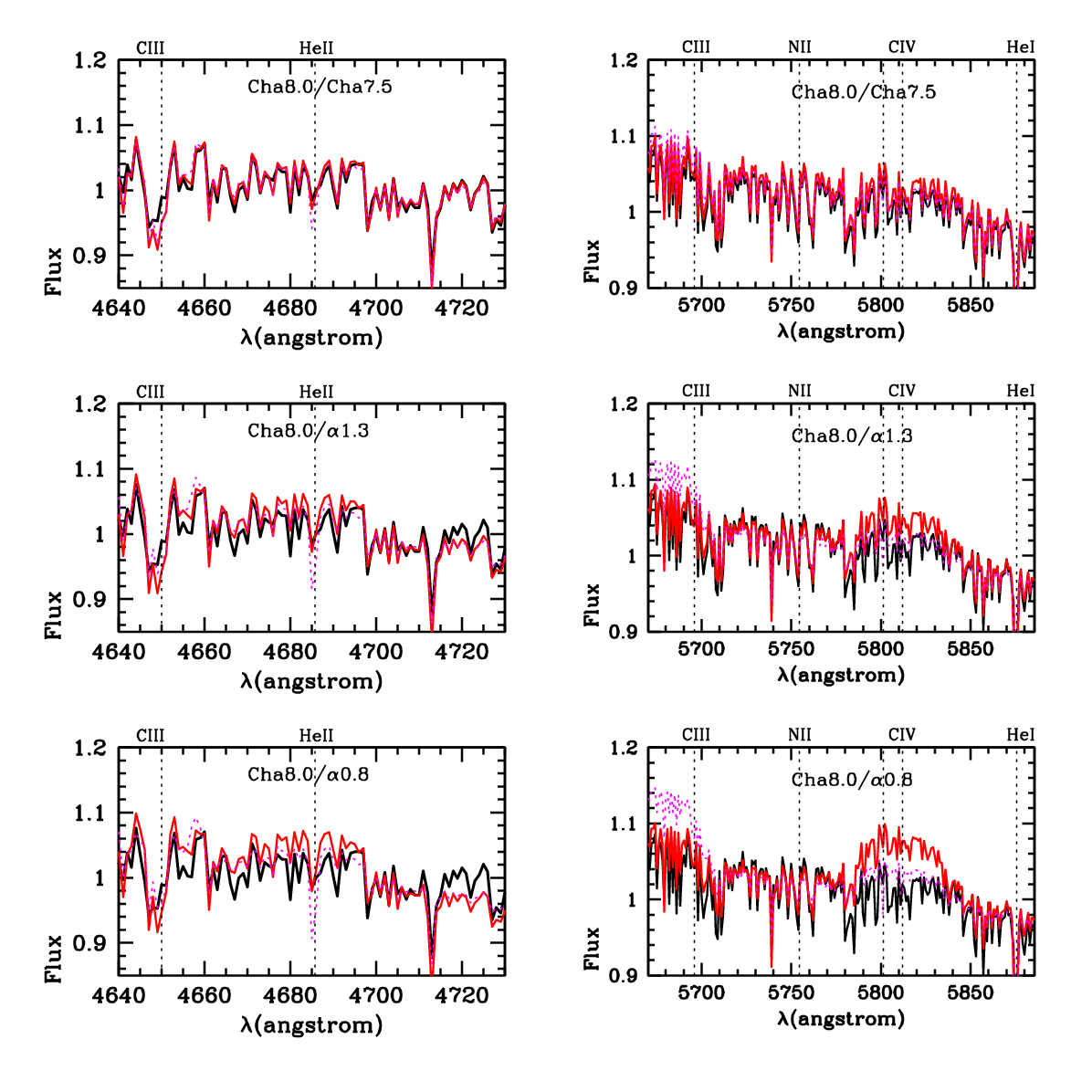}
\caption{ The reference model spectrum generated using HR-pyPopStar   
(Cha8.0)is compared with a variety of other model spectra 
plotted as coloured lines. In the top panels the coloured lines
are for constant star formation rate lasting $10^{7.5}$ years
with a Chabrier IMF. Dotted magenta lines are for the dust-free
case, while solid red lines include the effects of dust on the youngest
stars as explained in the text. The middle and bottom panels
are for models where the IMF has slope $\alpha=1.3$ and $\alpha=0.8$
above 10$M_{\odot}$. All models for solar metallicity.  
\label{models}}
\end{figure*}
\begin{figure*}
\includegraphics[width=101mm]{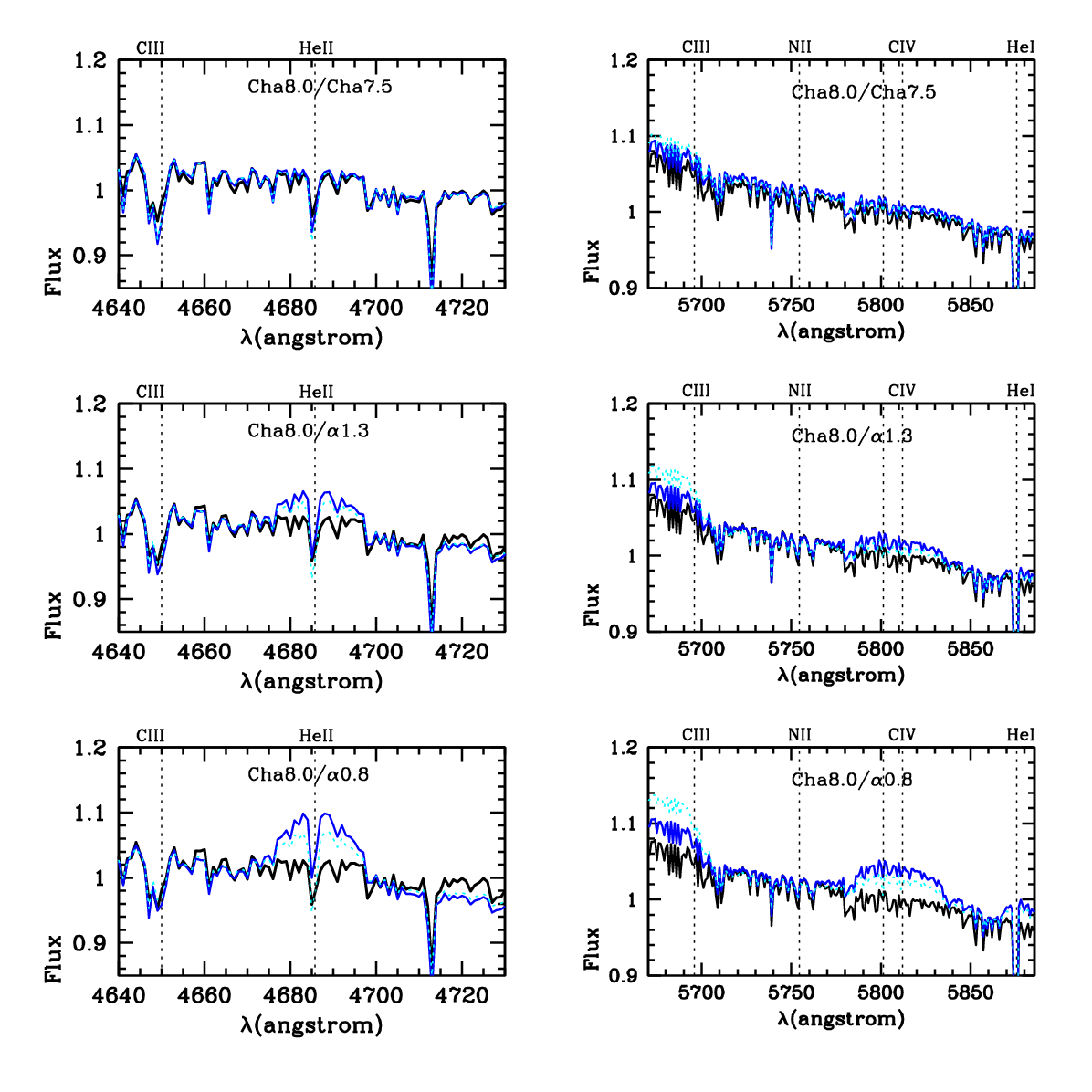}
\caption{ as in the previous figure, except at half solar metallicity.
\label{models}}
\end{figure*}

\section{Summary and final discussion}

We have extracted a sample of 17 nearby face-on galaxies from the final data
release of the SDSS-IV  MaNGA surveys where there are 100 or more spaxel
resolution elements within the half-light radius and more than a dozen
within 0.15$R_{50}$. The galaxies are selected by applying the condition
that the average H$\alpha$ equivalent width (corrected for dust extinction)
is 700 or more in at least one radial bin of width 0.1$R_{50}$. A value of
700 is chosen because it cannot be explained by a starburst of any age if the
stellar initial mass function has a Salpeter slope at the high mass end.
Spaxels with EQW(H$\alpha$)$>700$ define so-called H$\alpha$ excess regions
within the galaxies. Spaxels with EQW(H$\alpha$)$\sim$100 define normal
starburst regions.

The goal of our study is to ascertain whether the  H$\alpha$ excess regions
have an excess population of massive OB and Wolf-Rayet stars, to study
the interstellar medium conditions in these areas of the galaxy,
and to investigate whether there is evidence for accreting black holes
in addition to stellar sources of ionization. We now summarize our main findings,
dividing the discussion into three  topics.

\vspace{0.5cm}
{\bf Wolf-Rayet star features}
\begin{itemize}
\item Selecting spaxels for stacking in the two dimensional plane of H$\alpha$
equivalent width and an emission line ratio  that is insensitive
to dust, is more effective at revealing Wolf-Rayet features that selecting on
H$\alpha$ alone. We have tested two diagnostics of ionization parameter:
[OIII]/H$\beta$ and [SIII]/[SII] and find that the second  is more predictive
in  half of the galaxies.    
\item  Eight out of the 17 galaxies have detections of excess WR features
in a stacked spectrum. Detections  
are not defined at a threshold equivalent width of a particular line, but
by looking whether there is excess emission centered at either 4686 \AA\
(the expected centroid of HeII) or 5801 \AA\ (the expected centroid of CIV)
with respect to a "control" stacked spectrum of spaxels from normal
starburst regions of the same galaxy.    
\item Six out of the eight galaxies with excess WR features
have  excesses both at 4686 and 5801 \AA\, i.e. there is evidence for
the presence of both WN and WC stars.
\item The regions of the galaxies with Wolf-Rayet features range in size from
300 pc to 1.5 kpc in diameter.    
\item They are usually located at or near the center of the galaxy, but
the WR region is often covers a elongated, cone-like  region that is 
asymmetrically distributed with 
respect to the  centre of the galaxy defined in optical broad-band emission.
\end{itemize}

\vspace{0.5cm}
{\bf Associated high-ionization emission line regions} 
\begin{itemize}
\item
{\em Only} the H$\alpha$ excess spaxel regions where Wolf-Rayet features
are detected exhibit strong high ionization lines. Excess [NeIII] is detected in 4 out of 8
of the WR regions and there are tentative [FeX]  detections in 2 galaxies.
\end{itemize}

\vspace{0.5cm}
{\bf Local interstellar medium conditions} 
\begin{itemize}
\item
We use the NaI $\lambda\lambda$5890,5896 absorption line  doublet as a probe of the cool
neutral HI phase of the interstellar medium. The H$\alpha$ excess regions are
almost always associated with higher equivalent width NaI absorption lines than
the normal starburst regions. Significant velocity offsets of these features are 
not observed.
\item
We use the difference in EQW NaI between the H$\alpha$ excess regions and the normal starburst regions as
a probe of the cool gas compression in these regions. We find a larger gas compression factor
in the regions where Wolf Rayet features are also detected.  
\item The FWHM and the centroid shift  of the  [SIII]$\lambda$9531 emission line is used to probe physical
conditions in the ionized gas. FWHM line widths greater than $\sim$ 200 km/s are only obseved
for the regions containing Wolf-Rayet stars. The [SIII] line is always blue-shifted, indicating that
the gas is outflowing. The [SIII] equivalent width correlated with the continuum reddening suggesting that
the ionized, outflowing gas is carrying dust with it.   
\end{itemize}

There are two main future challenges: 1) to constrain the nature of the young stellar populations in the
central regions of the galaxies, and 2) to ascertain whether the high-ionization emission lines
originate from matter accreting onto black holes, from radio-emitting jets
or from from shocks produced by outflows from the central starbursts.

Further progress on (1) will require joint modelling of emission lines in HII regions and  Wolf-Rayet
star features. G\"otberg et al (2019) investigate the impact of stars stripped in binaries on the 
integrated spectra of stellar populations, focusing specifically on the emission rates of HI-, HeI-, and HeII-
ionizing photons assuming single-age bursts and continuous rates of star formation. They compare their
results to results from the Starburst99 models  which
do not include binaries, as well as  the BPASS models 
(Stanway \& Eldridge 2018), another population synthesis code that does include binaries.

The main result (see the right panel of Figure 4 in the paper) is that in a single-age starburst, 
stripped stars dominate the HI and HeI/HeII ionizing budget at all ages greater than 15-20 Myr.
In a system with continuous star formation, which we have argued is appropriate for the galaxies
studied in this paper, stripped stars contribute less than 10\% of the HI-ionizing budget and 
less than half the HeI-ionizing budget at all ages when comnpared to Starburst99 models.
In continuous models, stripped stars dominate 
the HeII ionizing budget at ages greater than 10 Myr, similar to the single-age burst case. 
These results suggest that if the AGN and
starburst contributions to the ionizing spectrum can be appropriately disentangled and the
effects of dust properly accounted for, the combination of  emission line and Wolf-Rayet star 
diagnostics in the integrated spectra of galaxies should be highly constraining.

Near-infrared IFU spectroscopy is very likely needed to provide a check on the results.
The near-IR spectrum of the Circinius galaxy (Maiolino et al 1997) reveals
clear HeI and Br$\gamma$ emission, and a weak HeII line detection 
in the wavelength range between 2.05 and 2.25 $\mu$m,
where dust extinction effects are much less severe. 
Within the central 12 pc the starburst age is estimated to be 
about $7 \times 10^7$ yrs, somewhat beyond the peak of the Wolf Rayet phase,
so it is likely that these features would be stronger in some of our more
extreme systems such as 8311.6104.

Future progress on (2) relies on high-resolution imaging. The advent of a new era in large telescopes
in space and advanced interferometric instrumentation on the ground will no doubt prove very
interesting in the coming years.

\vspace{1cm}
\noindent
{\bf Data Availability}\\
The data underlying this study are publically available at https://www.sdss4.org/dr17/manga/.
Additional data products generated for the study, such as the stacked spectra, will
be made available upon reasonable request to the corresponding author.

\vspace{1cm}
\noindent
{\bf Acknowledgments}\\

We thank Tim Heckman for helpful comments that helped
improve the analysis.

Funding for the Sloan Digital Sky 
Survey IV has been provided by the 
Alfred P. Sloan Foundation, the U.S. 
Department of Energy Office of 
Science, and the Participating 
Institutions. 

SDSS-IV acknowledges support and 
resources from the Center for High 
Performance Computing  at the 
University of Utah. The SDSS 
website is www.sdss4.org.

SDSS-IV is managed by the 
Astrophysical Research Consortium 
for the Participating Institutions 
of the SDSS Collaboration including 
the Brazilian Participation Group, 
the Carnegie Institution for Science, 
Carnegie Mellon University, Center for 
Astrophysics | Harvard \& 
Smithsonian, the Chilean Participation 
Group, the French Participation Group, 
Instituto de Astrof\'isica de 
Canarias, The Johns Hopkins 
University, Kavli Institute for the 
Physics and Mathematics of the 
Universe (IPMU) / University of 
Tokyo, the Korean Participation Group, 
Lawrence Berkeley National Laboratory, 
Leibniz Institut f\"ur Astrophysik 
Potsdam (AIP),  Max-Planck-Institut 
f\"ur Astronomie (MPIA Heidelberg), 
Max-Planck-Institut f\"ur 
Astrophysik (MPA Garching), 
Max-Planck-Institut f\"ur 
Extraterrestrische Physik (MPE), 
National Astronomical Observatories of 
China, New Mexico State University, 
New York University, University of 
Notre Dame, Observat\'ario 
Nacional / MCTI, The Ohio State 
University, Pennsylvania State 
University, Shanghai 
Astronomical Observatory, United 
Kingdom Participation Group, 
Universidad Nacional Aut\'onoma 
de M\'exico, University of Arizona, 
University of Colorado Boulder, 
University of Oxford, University of 
Portsmouth, University of Utah, 
University of Virginia, University 
of Washington, University of 
Wisconsin, Vanderbilt University, 
and Yale University.




\appendix

\section {Additional maps of physical quantities}

In Figures A1 and A2, we show maps of the Balmer absorption line feature H8. We select this high-order
line because it is never contaminated by emission and easily measured in all the spectra. To zero'th 
order it provides a measure of the age  of the underlying stellar population, with
strong absorption indicating a young stellar population and weak emission an older stellar
population. The strength of the Balmer absorption feature peaks in stars of spectral types A-F and
is weaker in O and B stars, so that in the very youngest starbursts, Balmer absorption lines such
as H8 may also be weak. As can be seen in the two figures, H8 is almost always strongest in the central
regions of the galaxies in both the WR-detected and non-detected subsamples. There is no clear
difference in H8 values between the two sub-samples. This is  expected, because the  Wolf-Rayet
phase is short-lived  -- The WR phase occurs $ \sim 10^7$ years after the
initial starburst,  compared to $10^9$ years for the H8 absorption line
to reach maximum strength. 

Figures A3 and A4  show the ionized gas velocity maps for the galaxies in the two subsamples.
The velocities are measured from the shift of the centroid of the H$\alpha$ emission line with
respect to the systemic velocity. Each colour indicates a interval of 50 km/s in velocity.
If we define galaxies to have a significant velocity gradient if there is a clear gradient spanning
200 km/s in total (i.e. clear regions with 4 different coulours in the plots), we see that
five out of eight galaxies have significant velocity gradients in the subsample with WR
detections and four out of six in the subsample withour WR detections. One galaxy in the
WR-detected subsample exhibits strongly red-shifted emission across the enire region
within R$_{50}$ and two galaxies exhibit very weak velocity gradients. In the subsample with 
no WR detection, one galaxy has strong blue-shifted emission across most of the 
region within  R$_{50}$ and one galaxy has a very weak velocity gradient. In conclusion, there
is no clear difference in the kinematics of the ionized gas in the two sub-samples. 

\begin{figure*}
\includegraphics[width=121mm]{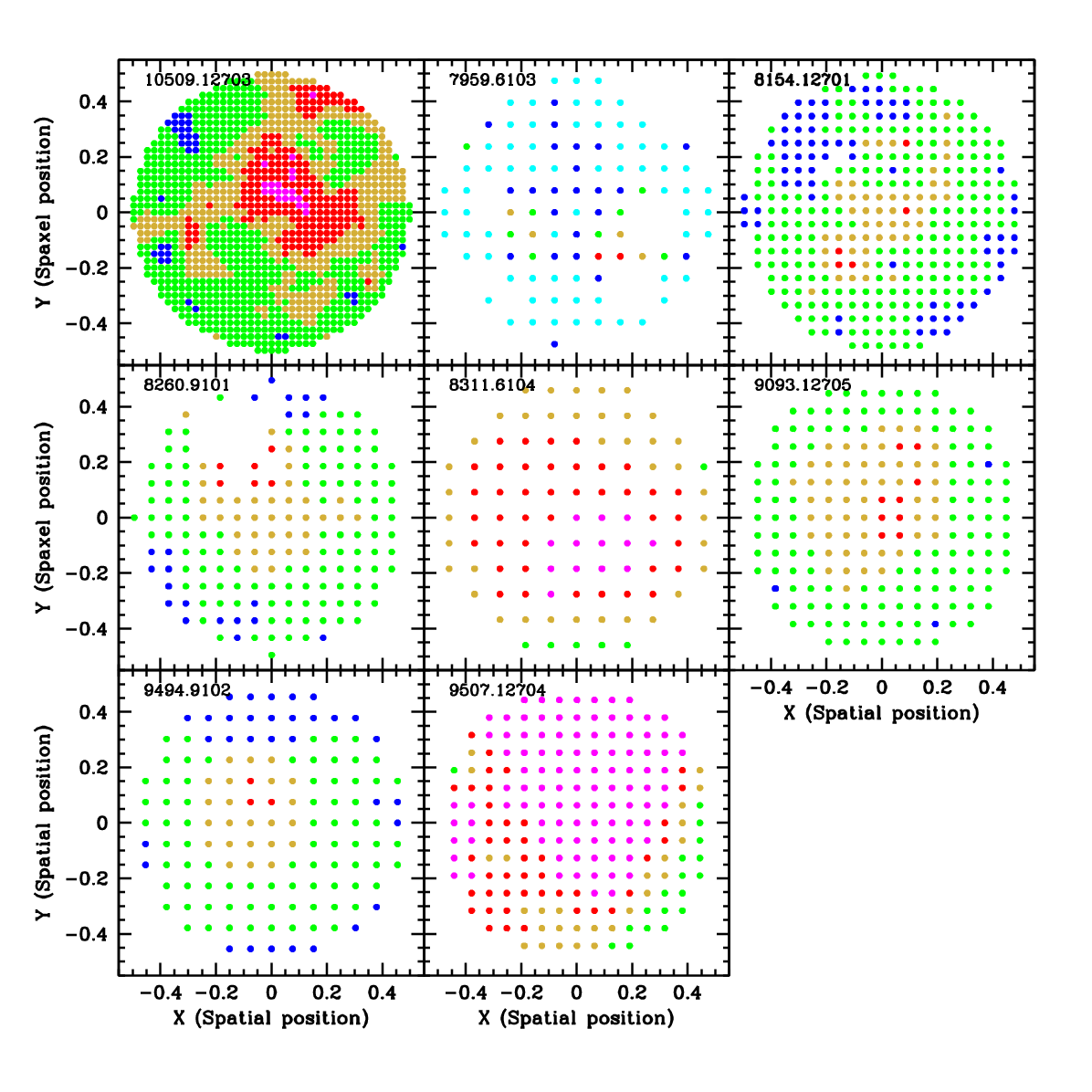}
\caption{Two-dimensional maps of the Balmer absorption line index H8  for the 8 galaxies
with Wolf-Rayet detections. The circular points indicate each spaxel measurement and are
colour-coded as follows: magenta (H8 $>$0.8), red (0.6$<$ H8 $<$0.8),  
dark gold (0.5$<$ H8 $<$0.6), green (0.4$<$ H8 $<$0.5),
blue  (0.3$<$ H8 $<$0.4), cyan (0.2$<$ H8 $<$0.3).
\label{models}}
\end{figure*}
\begin{figure*}
\includegraphics[width=121mm]{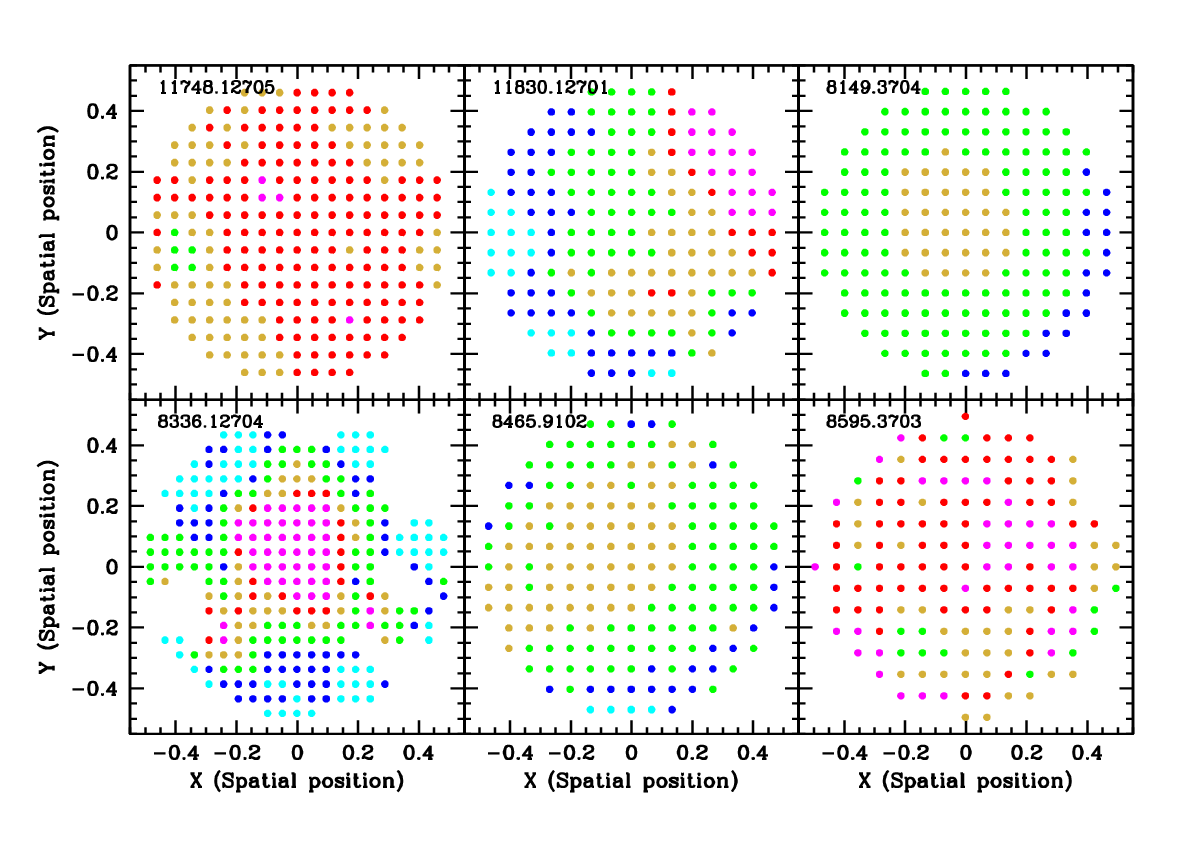}
\caption{As in the previous figure, except for the 6 galaxies with no Wolf Rayet 
detections. 
\label{models}}
\end{figure*}

\begin{figure*}
\includegraphics[width=121mm]{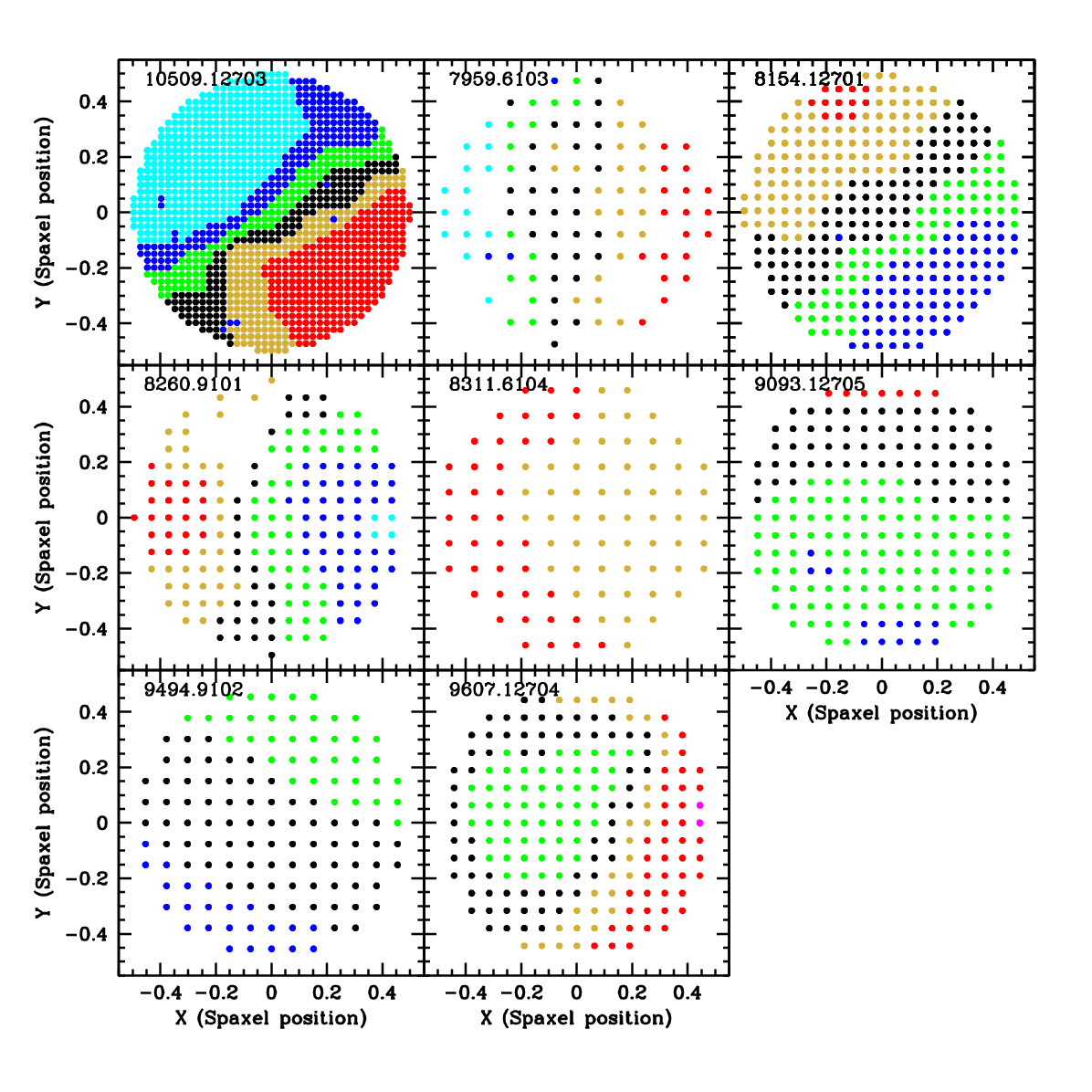}
\caption{Two-dimensional maps of the velocity of the H$\alpha$ emission line   for the 8 galaxies
with Wolf-Rayet detections. The circular points indicate each spaxel measurement and are
colour-coded as follows: black (0$<$V(H$\alpha$)$<$50 km/s ), green (0$<$V(H$\alpha$)$>$-50 km/s),    
dark gold (50$<$V(H$\alpha$)$<$100 km/s), blue (-50$<$V(H$\alpha$)$>$-100 km/s),    
red (100$<$V(H$\alpha$)$<$200 km/s), cyan (-100$<$V(H$\alpha$)$>$-200 km/s),    
magenta(300 km/s $<$V(H$\alpha$), cyan bold ((-300 km/s $>$V(H$\alpha$).
\label{models}}
\end{figure*}

\begin{figure*}
\includegraphics[width=121mm]{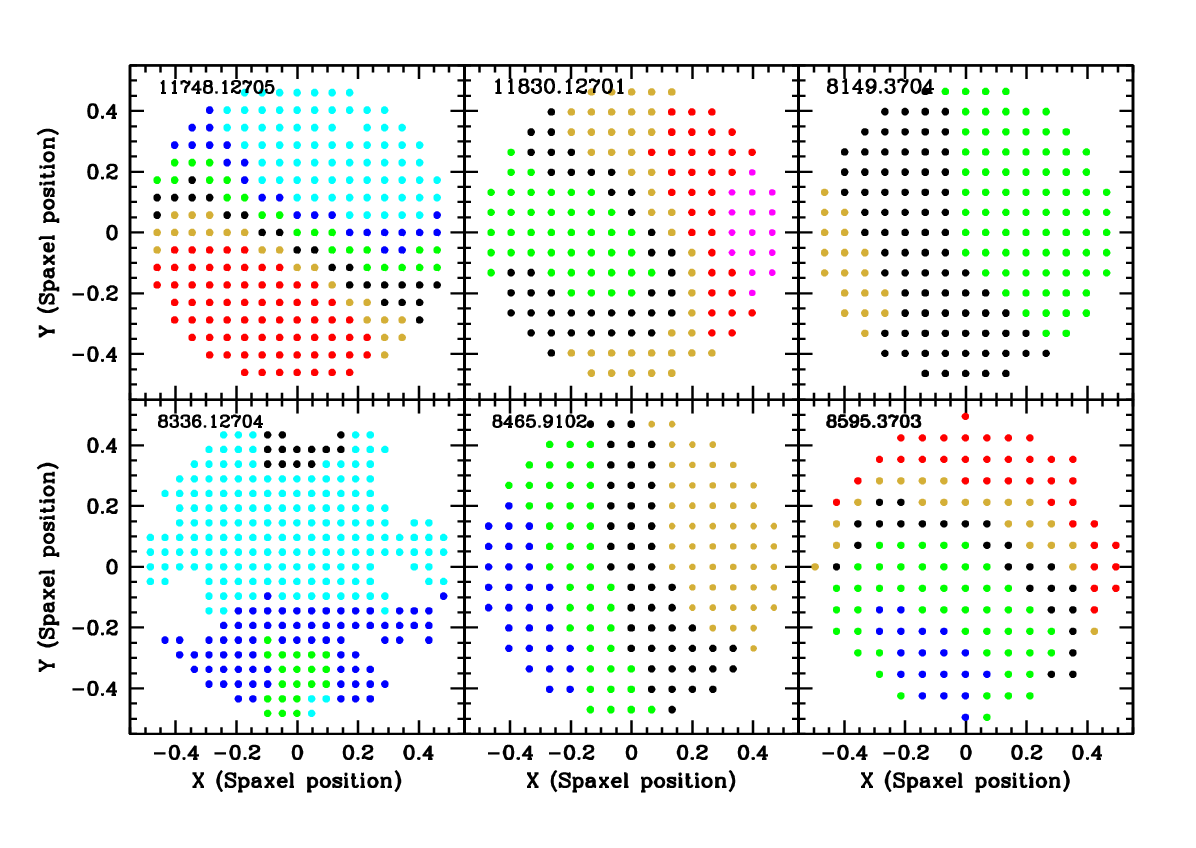}
\caption{As in the previous figure, except for the 6 galaxies with no Wolf Rayet 
detections. 
\label{models}}
\end{figure*}

\section {The blue and red bump regions of the sub-sample with no Wolf Rayet detections}

Figures B1 and B2 show stacked spectra in the blue and red bump wavelength regions for the sub-sample listed in
Table 1 as having no excess WR feature detections. As in Figures 9 and 10, the stacked spectra of the
regions with normal H$\alpha$ equivalent width values are plotted in back, while the stacked
spectra of the  H$\alpha$ excess regions are plotted in red. The 1$\sigma$ error region on the  H$\alpha$ excess
stacked spectra are plotted in cyan.  With the possible exception of object 8149.3704, no excess 
emission is seen close to the HeII$\lambda$4686  or the CIV$\lambda$5801 lines.  

\begin{figure*}
\includegraphics[width=105mm,angle=270]{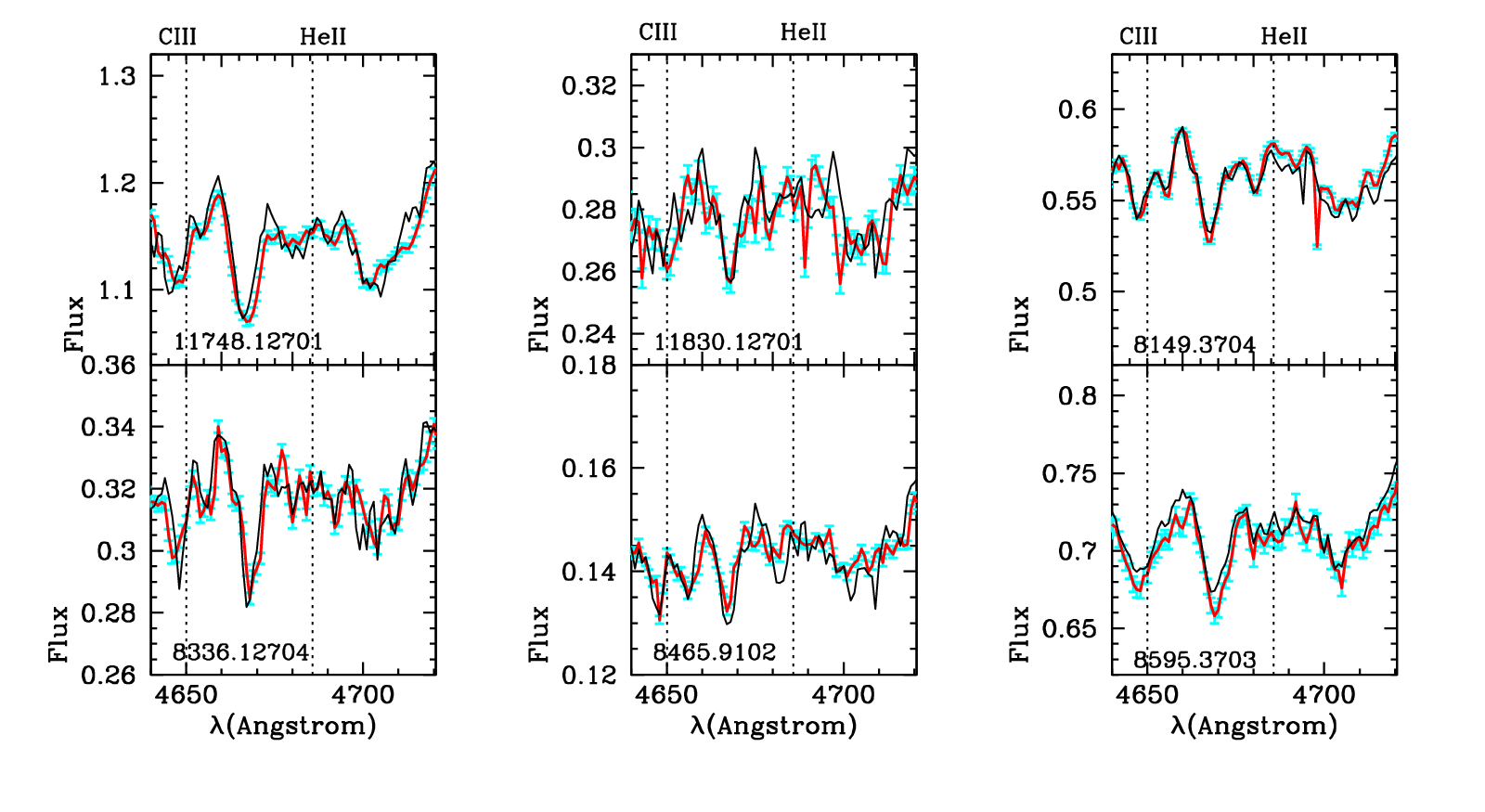}
\caption{The ``blue bump'' region of the stacked spectra is shown  for the selected H$\alpha$ excess
spaxels (red or magenta lines) compared to the spaxels with normal starburst values of the   H$\alpha$ equivalent
(black lines). Errorbars on the  H$\alpha$ excess spectra are plotted in cyan.
All spectra are selected using a cut on $\log$ [OIII]/H$\beta$ (the attempt to select using
sulphur line ratios did not result in better detections that shown.  
\label{models}}
\end{figure*}
\begin{figure*}
\includegraphics[width=105mm,angle=270]{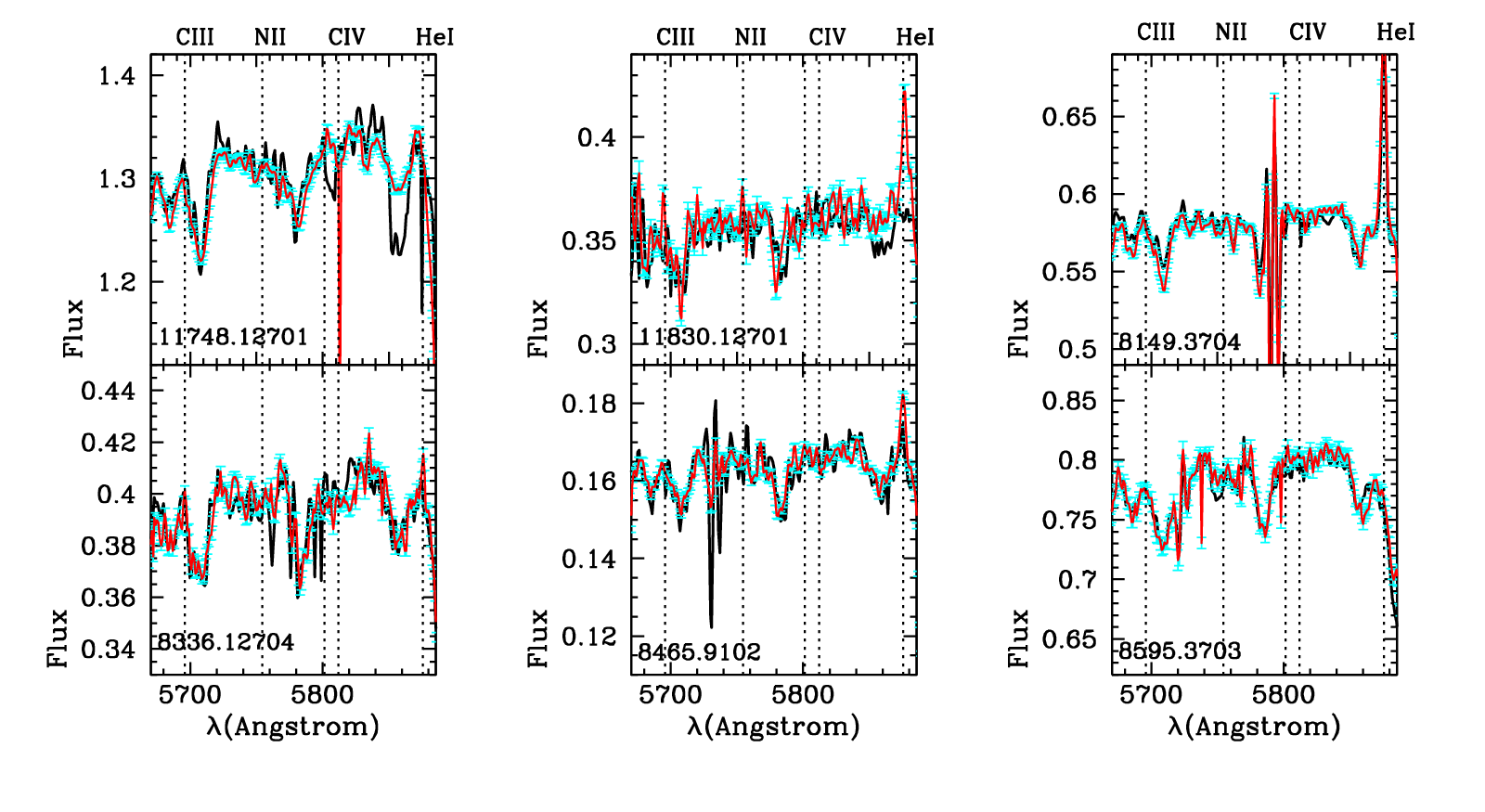}
\caption{As in the previous figure, except for the ``red bump'' region of the stacked spectra.                                                    
\label{models}}
\end{figure*}

\end{document}